\documentclass[aps,floatfix,10pt,twocolumn,amsmath,amssymb,superscriptaddress,nobibnotes,prb,
longbibliography]{revtex4-2}
\usepackage{graphicx}
\usepackage{enumerate}
\usepackage[colorlinks=true,linktoc=all,linkcolor=blue,breaklinks=true,citecolor=blue,urlcolor=blue]{hyperref}
\usepackage{textcomp}
\usepackage{amsmath}
\usepackage{amssymb}
\usepackage{pgf}
\usepackage{subfigure}
\usepackage[english]{babel}
\usepackage[normalem]{ulem}
\usepackage{tabularray}
\usepackage{xcolor}
\usepackage{array}
\usepackage{soul}

\newcommand{\ket}[1]{|#1\rangle}

\newcommand{\lambdap}{{\rm g}}
\newcommand{\lambdam}{{\rm u}}
\newcommand{\lambdapm}{{\rm g/u}}
\newcommand{\lambdamp}{{\rm u/g}}
\newcommand{\qubitp}{0}
\newcommand{\qubitm}{1}
\newcommand{\qubitpm}{0,1}

\begin{document}
	
	\title{Fermion-parity qubit in a proximitized double quantum dot}
	\author{Max Geier}
	\affiliation{Center for Quantum Devices, Niels Bohr Institute, University of Copenhagen, 2100 Copenhagen, Denmark}
        \affiliation{Department of Physics, Massachusetts Institute of Technology, Cambridge, MA 02139, USA}
	
	\author{Rub\'en Seoane Souto}
	\affiliation{Center for Quantum Devices, Niels Bohr Institute, University of Copenhagen, 2100 Copenhagen, Denmark}
	\affiliation{Division of Solid State Physics and NanoLund, Lund University, 22100 Lund, Sweden}
	\affiliation{Departamento de Física Teórica de la Materia Condensada, Condensed Matter Physics Center (IFIMAC) and Instituto Nicolás Cabrera, Universidad Autónoma de Madrid, 28049 Madrid, Spain}
	\affiliation{Instituto de Ciencia de Materiales de Madrid (ICMM), Consejo Superior de Investigaciones Científicas (CSIC),
		Sor Juana Inés de la Cruz 3, 28049 Madrid, Spain.}
	
	\author{Jens Schulenborg}
	\affiliation{Center for Quantum Devices, Niels Bohr Institute, University of Copenhagen, 2100 Copenhagen, Denmark}
	\affiliation{Department of Microtechnology and Nanoscience (MC2), Chalmers University of Technology, S-412 96 G\"oteborg, Sweden}
	
	\author{Serwan Asaad}
	\affiliation{Center for Quantum Devices, Niels Bohr Institute, University of Copenhagen, 2100 Copenhagen, Denmark}
	
	\author{Martin Leijnse}
	\affiliation{Center for Quantum Devices, Niels Bohr Institute, University of Copenhagen, 2100 Copenhagen, Denmark}
	\affiliation{Division of Solid State Physics and NanoLund, Lund University, 22100 Lund, Sweden}
	
	\author{Karsten Flensberg}
	\affiliation{Center for Quantum Devices, Niels Bohr Institute, University of Copenhagen, 2100 Copenhagen, Denmark}
	
	\date{\today}
	
	\begin{abstract}
		Bound states in quantum dots coupled to superconductors can be in a coherent superposition of states with different electron number but with the same fermion parity. Electrostatic gating can tune this superposition to a sweet spot, 
		where the quantum dot has the same mean electric charge independent of its electron-number parity. Here, we propose to encode quantum information in the local fermion parity of two tunnel-coupled quantum dots embedded in a Josephson junction. At the sweet spot, the qubit states have zero charge dipole moment. 
        This protects the qubit from dephasing due to charge noise acting on the potential of each dot, as well as fluctuations of the (weak) inter-dot tunneling. At weak inter-dot tunneling, relaxation is suppressed because of disjoint qubit states. On the other hand, for strong inter-dot tunneling the system is protected against noise affecting each quantum dot separately (energy level noise, dot-superconductor tunneling fluctuations, and hyperfine interactions). Finally, we describe initialization and readout as well as single-qubit and two-qubit gates by pulsing gate voltages.
	\end{abstract}
 
	\maketitle
	
\section{Introduction}
	\label{sec:intro}
	
	Quantum dots coupled to superconductors host bound states with energies below the superconducting gap.  They are known as Yu-Shiba-Rusinov states \cite{YuActaPhysSin2005Aug, ShibaProgTheorPhys1968Sep, Rusinov1969Jan} for large charging energy or Andreev bound states \cite{Kulik1969} with small charging energy compared to the superconducting gap. These bound states are superpositions with different particle number due to so-called Andreev tunnel events where pairs of electrons in the quantum dot are transferred as a Cooper pair in the superconductor. This process thus preserves the total fermion parity of the system. In recent years, hybrid superconductor-semiconductor structures have proven to be a reliable platform to realize Yu-Shiba-Rusinov \cite{HatterNatCommun2015Nov,JellinggaardPhysRevB2016Aug,Grove-RasmussenNatCommun2018Jun} and Andreev bound states \cite{HofstetterNature2009Oct,DeaconPhysRevLett2010Feb,PilletNatPhys2010Dec,LeeNatNanotechnol2014Jan,SuNatCommun2017Sep,ShenNatCommun2018Nov,PoschlPhysRevB2022Oct,BanerjeePhysRevLett2023Mar}. In Josephson junctions, these states can induce $0-\pi$ transitions \cite{Bulaevskii1977Apr,RozhkovPhysRevLett1999Mar, VecinoPhysRevB2003Jul, vanDamNature2006Aug, Martin-RoderoAdvPhys2011Dec, ZondaSciRep2015Mar, DelagrangePhysRevB2016May,WhiticarPhysRevB2021Jun, RazmadzePhysRevB2023Feb}, $\phi_0$ phase shifts \cite{PilletNanoLett2019Oct,KurtossyNanoLett2021Oct}, occupation \cite{EstradaSaldanaPhysRevLett2018Dec} and spin-dependent \cite{BoumanPhysRevB2020Dec} transport, and qubits \cite{Janvier_Science2015,Hays_PRL2018,Hays_Science2021,PitaVidalNatPhys2023Aug}. Hybridized Andreev bound states have been applied as Cooper pair splitters \cite{WangNature2022Dec} and to create minimal Kitaev chains \cite{LeijnseFlensberg2012,SauNatCommun2012Jul,TomDvirNature2023Feb}.

 	Electrostatic gating can control the mean electric charge of the subgap states \cite{BauerJPhys:CondensMatter2007Nov, MengPhysRevB2009Jun,LeeNatNanotechnol2014Jan,SchindelePhysRevB2014Jan,Janvier_Science2015,Hays_PRL2018,PoschlPhysRevB2022Dec}. The quantum dot can be tuned to a sweet spot where it has same mean electric charge for both ground states with an even or odd electron number parity. As a consequence, both fermion parity sectors have the same response to small electric fields, and the energy difference between the two sectors are thus protected against small fluctuations of the electrostatic environment.
	
	\begin{figure}[b!]
		\includegraphics[width=1\linewidth]{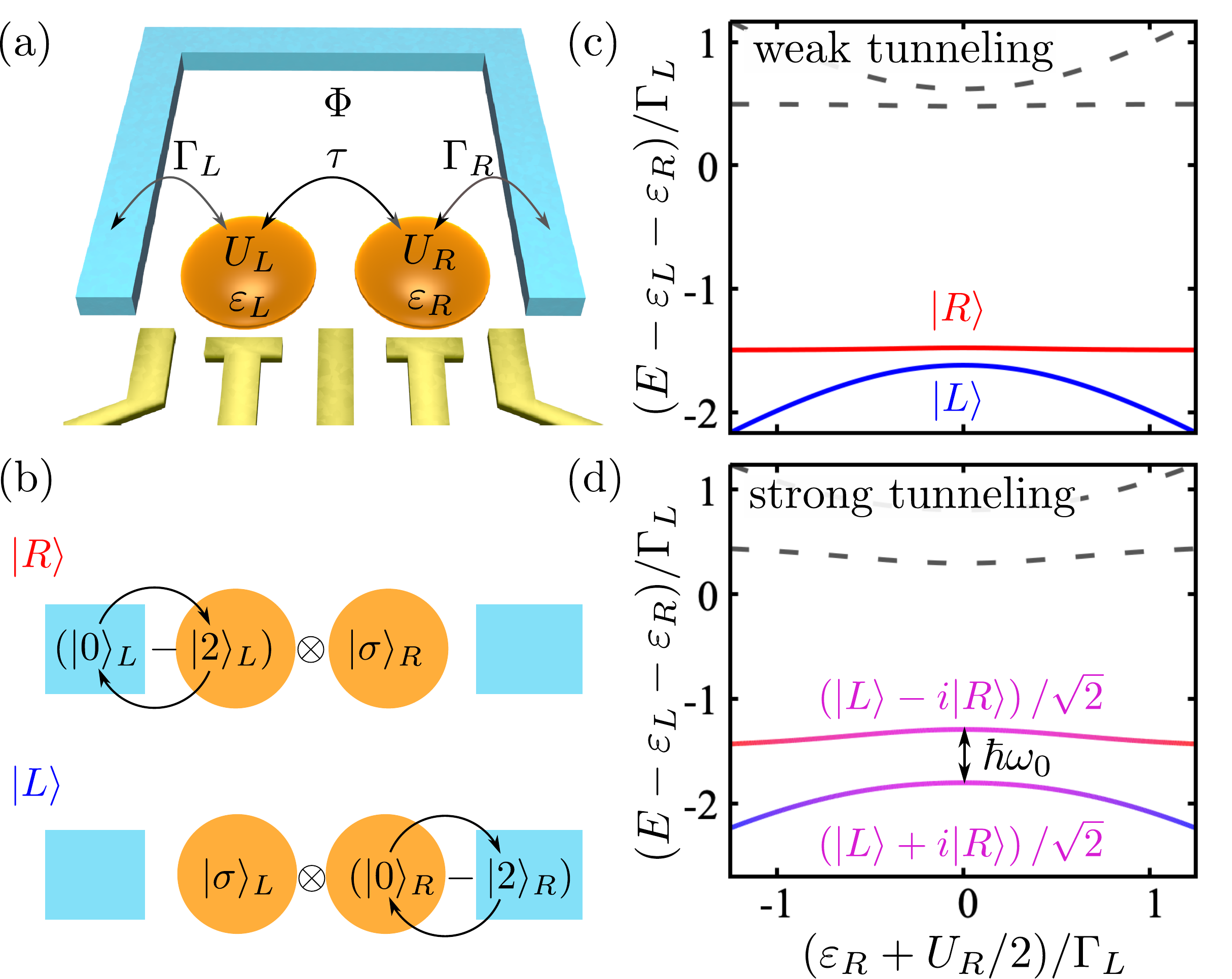}
		\caption{\label{fig:1}
			(a) Device sketch: Two quantum dots labeled by $\nu = L, R$ (orange) with charging energy $U_\nu$ are coupled to a superconducting loop (light blue) that induces superconducting correlations with magnitude $\Gamma_{\nu}$ and phase difference $\phi = 2\pi \Phi/\Phi_0$ controlled by the enclosed magnetic flux $\Phi$, where $\Phi_0 = h/2e$ is the superconducting flux quantum. The dot level energies $\varepsilon_{\nu}$ and interdot coupling $\tau$ are tunable by gate electrodes (yellow).
			(b) Sketch of the states $|L\rangle$ and $|R\rangle$ whose superposition makes up the qubit eigen states. The states $|L\rangle$ and $|R\rangle$ are a product of single electrons of spin $\sigma$, $\ket{\sigma}_\nu=\hat{c}^{\dagger}_{\nu\sigma}\ket{0}_\nu$, and even parity states in an equal superposition of empty $\ket{0}_\nu$ and double occupation $\ket{2}_\nu=\hat{c}^{\dagger}_{\nu\uparrow}\hat{c}^{\dagger}_{\nu\downarrow}\ket{0}_\nu$.
			(c) Many-body spectrum as a function of level energy $\varepsilon_{R}+U_R/2$ for weak tunneling.
			The blue (red) solid line indicates the qubit state $|L\rangle$ ($|R\rangle$), while dashed lines are excited states. Parameters: $\tau = \Gamma_L/20,\ \Gamma_R = 1.1\Gamma_L,\  U_L = U_R = \Gamma_L$, $\varepsilon_L = -U_L/2$. 
			(d) Corresponding results for strong tunneling.
			The line color indicates the wave function overlap with $|L\rangle$ (blue) and $|R\rangle$ (red). 
                The qubit eigen energy at the operating point is $\hbar \omega_0$.
			Parameters: $\tau = \Gamma_L/4$, remaining parameters unchanged.
		}
	\end{figure}
	
Here, we propose to leverage the protection of the local fermion parity together with the tunability of the charge expectation value to define a qubit in a pair of electrostatically controlled quantum dots embedded in a superconducting loop, see sketch in Fig.~\ref{fig:1}(a). The two qubit states can be encoded in the protected local fermion parity of the two dots: the state $|L\rangle$ ($|R\rangle$) is a product of the left (right) quantum dot hosting an odd number of fermions while the right (left) quantum dot hosts an even number, as depicted in Fig.~\ref{fig:1}(b). 
The quantum dot with odd fermion parity has a spin $1/2$ degree of freedom. When the coupling of the spin to the environment is negligible, the qubit can be operated with spin-degenerate levels. Otherwise, the spin can be polarized by an applied magnetic field.

The coherence properties of this system depend on the hybridization between the quantum dots. For weak hybridization relative to the energy difference of the left and right bound states, the qubit eigen states are $|L\rangle$ and $|R\rangle$. Relaxation of the qubit state requires a quasiparticle to tunnel between the quantum dots, which is suppressed for weak tunneling. Electric field fluctuations enter only to second order in the qubit spectrum by fluctuations of the inter-dot tunneling strength and the level energy. For strong hybridization, the qubit eigen states are close to an equal superposition of $|L\rangle$ and $|R\rangle$.
In this regime, dephasing due to noise coupling to individual quantum dots (such as local magnetic field or level energy fluctuations) are suppressed. Instead, the qubit is sensitive to fluctuations in the tunneling strength between the quantum dots.In both regimes, the qubit is insensitive towards common-mode fluctuations in parameters coupling to individual quantum dots.

It is instructive to consider the case where the charging energy can be neglected. In the case, the qubit states can be described in terms of a single Bogoliubov quasiparticle shared between the quantum dots. This quasiparticle has zero electric charge when the quantum dots are tuned to the sweet spot. From this point of view, our proposal can be considered a ``chargeless'' variant of a charge qubit \cite{HayashiPhysRevLett2003Nov, FujisawaPhysicaE2004Mar}. In our parity qubit, the superconducting correlations strip off the electron's charge.

The fermion parity qubit combines features of single-electron charge \cite{HayashiPhysRevLett2003Nov, FujisawaPhysicaE2004Mar,PeterssonPhysRevLett2010Dec} and spin qubits \cite{BurkardRevModPhys2023Jun} with features of superconducting platforms \cite{KrantzAPR2019,KjaergaardAnnuRevCondensMatterPhys2020Mar}.  In comparison to single-electron charge and spin qubits, (i) the fermion-parity qubit provides intrinsic protection from electric field noise due to the chargeless nature of the quasiparticles, 
(ii) can be operated with spin-polarized qubit states, (iii) is directly electrically addressable, (iv) can provide protection from relaxation by spatially separating the two qubit states in the weak-tunneling limit, and (v) is compatible with superconducting circuit architecture. In comparison to superconducting qubits, such as transmons \cite{KochPhysRevA2007Oct}, the fermion parity qubit allows for faster gate times because the qubit subspace can be well separated from the next excited states.

 In the following Sec.~\ref{sec:qubit}, we define the system Hamiltonian and the regimes for qubit operations, as well as microwave controlled single- and two-qubit rotations. A detailed discussion on the effects of noise on the qubit is contained in Sec.~\ref{sec:noise}. 
 While Sec.~\ref{sec:qubit} and \ref{sec:noise} describe the system in the limit of an infinite gap $\Delta$ in the superconducting leads, in Sec.~\ref{sec:largeU} we discuss the case when charging energy in the quantum dots is the largest energy scale.
 Sections~\ref{sec:gate_times} and \ref{sec:coherence_times} provide estimates of gate and coherence times, respectively, for realistic systems.
 In Sec.~\ref{sec:intro_qubits}, we discuss the relation of our proposal to other qubit realizations in detail.


\section{The fermion-parity qubit}
\label{sec:qubit}

In this section, we define the Hamiltonian describing the pair of quantum dots connected to superconductors (Sec.~\ref{sec:qubit_H}), identify the operating regimes (Sec.~\ref{sec:qubit_regimes}), and describe initialization and read-out (Sec.~\ref{sec:qubit_initialization}) as well as single- and two-qubit rotations (Sec.~\ref{sec:qubit_rot} and \ref{sec:qubit_twoqubitgates}).

\subsection{System Hamiltonian}
\label{sec:qubit_H}

Figure~\ref{fig:1}(a) shows a sketch of the fermion-parity qubit setup with the relevant control parameters. The system Hamiltonian,
\begin{equation}
\hat{H}=\sum_\nu \hat{H}_\nu+\hat{H}_T\,,
\label{eq:full_Hamiltonian}
\end{equation} 
is composed of terms $\hat{H}_\nu,\ \nu = L,R$ describing the two individual quantum dots and their tunnel coupling $\hat{H}_T$. The terms $\hat{H}_\nu$ describing the individual quantum dots read
\begin{equation}
\hat{H}_{\nu}=\sum_{\sigma}\varepsilon_{\nu}\hat{n}_{\sigma\nu}+U_\nu \hat{n}_{\uparrow\nu}\hat{n}_{\downarrow\nu}+\Gamma_\nu\left( \hat{c}_{\uparrow\nu}\hat{c}_{\downarrow\nu}+{\rm H.c.}\right)\,,
\label{eq:H_nu}
\end{equation}
where $\hat{n}_{\sigma\nu}=\hat{c}_{\sigma\nu}^\dagger \hat{c}_{\sigma\nu}$, with the annihilation operator $\hat{c}_{\sigma\nu}$ for an electron on the dot $\nu$ with spin $\sigma = \ \uparrow, \downarrow$. The level energy $\varepsilon_\nu$ can be controlled using electrostatic gates, $U_\nu$ is the Coulomb repulsion strength, and $\Gamma_\nu$ describes the proximity-induced superconducting correlations in the quantum dots.  
The proximity-induced pairing strength enters in second-order perturbation theory, $\Gamma_\nu = \pi t^2_{\nu} \nu_{\rm S}$, where $t_\nu$ is the hopping amplitude between quantum dot $\nu$ and superconductor and $\nu_{\rm S}$ is the density of states of the superconductor. We here consider only a single level $\varepsilon_\nu$ in each dot and thus each dot can only be occupied by up to two electrons. This description is valid in the limit of a large gap in the superconducting leads $\Delta \gg \Gamma_\nu, \varepsilon_\nu, U_\nu$. 
We focus on this limit throughout the bulk of the paper due to its analytical tractability and for presentation purposes. 
The qualitative qubit properties and operation principles also apply in the limit $U_\nu \gg \Delta, \Gamma_\nu$ as we discuss in Sec.~\ref{sec:largeU} based on numerical calculations.

The eigen states of $\hat{H}_\nu$ with even local fermion parity are superpositions of zero $|\rm{vac}\rangle_\nu$ and two excess electrons of opposite spin $|2\rangle_\nu = \hat{c}_{\nu \uparrow}^\dagger \hat{c}_{\nu \downarrow}^\dagger |\rm{vac} \rangle_\nu$. The odd local fermion parity subspace contains the states where a single electron of spin $\sigma$ occupies the quantum dot, $|\sigma\rangle_\nu = \hat{c}_{\nu \sigma}^\dagger |\rm{vac} \rangle_\nu$. 

The tunnel coupling between the dots is given by
\begin{equation}
\hat{H}_\tau=\tau\sum_\sigma e^{i\phi/2} \hat{c}^{\dagger}_{\sigma R}\hat{c}_{\sigma L} +{\rm H.c.}\,,
\label{eq:H_tau}
\end{equation}
where $\tau$ is the tunneling amplitude and $\phi$ the superconducting phase difference.
The Hamiltonian is written in a gauge where the pairing amplitudes $\Gamma_\nu$ are real and positive while the superconducting phase difference is included in the tunneling term. For now, spin-orbit coupling and Zeeman field are neglected as these terms are not required for operating the fermion-parity qubit. These effects are included in the discussion on dephasing and relaxation due to parameter fluctuations in Sec.~\ref{sec:noise}. 

\subsection{Sweet spot}
\label{sec:qubit_regimes}

Crucial to our proposal is the sweet spot at which decoherence is suppressed. Here, we describe how to reach this sweet spot, and how the main control parameters affect qubit operations at or close to this point.

{\em Level energies $\varepsilon_\nu$.---} The sweet spot, at which both fermion-parity sectors of a quantum dot have the same mean charge, is reached by setting the level energy to $\varepsilon_\nu=-U_\nu/2$. At this point, the quantum dot eigen states with even fermion parity are symmetric (labeled by $\lambda_\nu = \lambdap$) and antisymmetric (labeled by $\lambda_\nu = \lambdam$) superpositions $|\lambdapm\rangle_\nu = \frac{1}{\sqrt{2}}(|\rm{vac}\rangle_\nu \pm | 2\rangle_\nu)$ with energies $\pm \Gamma_\nu$. Thus the ground state of individual quantum dots is $|\lambdam\rangle_\nu$. 

\begin{figure}
    \centering
    \includegraphics[width=\columnwidth]{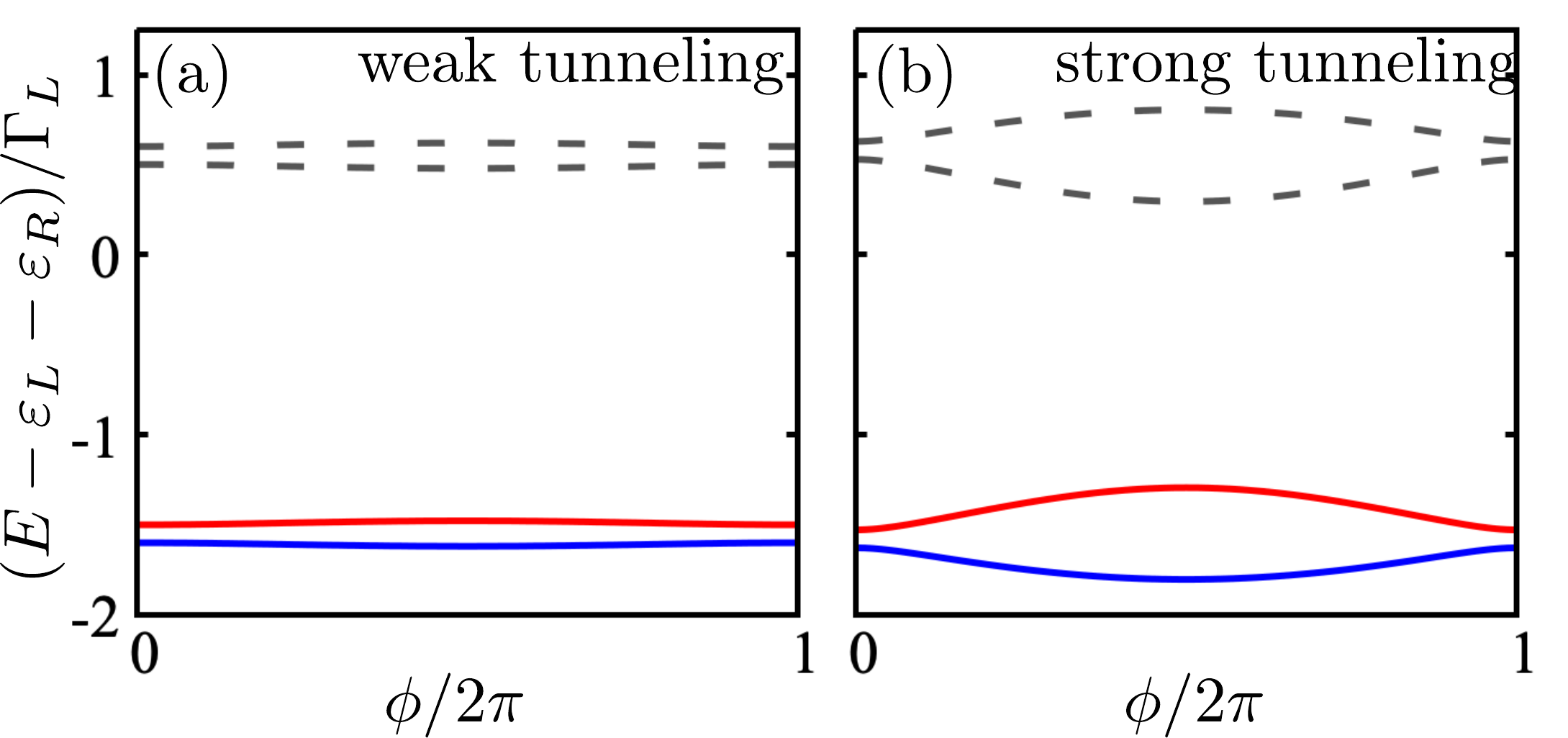}
    \caption{
    Dependence of the spectrum of Eq.~\eqref{eq:full_Hamiltonian} on the superconducting phase difference $\phi$ at the operating point $\varepsilon_\nu = - U_\nu / 2$ in (a) the weak tunneling regime $\tau = 0.05 \Gamma_L$ and (b) the strong tunneling regime $\tau = 0.25 \Gamma_L$. Remaining parameters as in Fig.~\ref{fig:1}.} 
    \label{fig:phase_depedence}
\end{figure}

{\em Superconducting phase difference $\phi$.---}
The phase difference $\phi$ determines the tunneling between the quantum dots. At $\phi = 0$, tunneling of an electron switches the symmetry of the even-parity bound state ($\lambda_\nu = \lambdam \leftrightarrow \lambdap$), due to a sign acquired by fermionic anticommutation upon tunneling. At $\phi = \pi$, this sign is cancelled and the symmetry of the even-parity bound state is preserved, thus coupling the two qubit states, which have $\lambda_R = \lambda_L = \lambdam$. 
This manifests in the spectra as a function of phase difference shown in Fig.~\ref{fig:phase_depedence}. 
At $\phi = \pi$ the gap in the spectrum is proportional to $\tau$. For $\phi = 0$, tunneling between the left and right quantum dots is prohibited and the gap in the spectrum is given by the asymmetry of the two quantum dots $|\langle L|H_L|L\rangle - \langle R|H_R|R\rangle| = |(U_R - U_L)/2 - (\Gamma_R - \Gamma_L)|$ at the sweet spot.
Setting $\phi=0$ provides an effective way to switch off the tunneling coupling and thus parking the qubit in the protected situation. In contrast, for qubit operations that involve rotating in the $|L\rangle,|R\rangle$ space, we set $\phi = \pi$.  Moreover, at $\phi = \pi$, the spectrum is first-order insensitive to the fluctuations in $\phi$ (see Fig.~\ref{fig:phase_depedence} and Eq.~\eqref{eq:omega01^(2)_phi}).

{\em Tunneling strength $\tau$.---}
Tunneling between the quantum dots hybridizes the bound states in the two dots. Without tunnel coupling, the qubit eigen states are the product states $|L\rangle = |\sigma\rangle_L \otimes |\lambdam\rangle_R$, $|R\rangle = |\lambdam\rangle_L \otimes |\sigma\rangle_R$ with energies $-U_{L/R}/2 - \Gamma_{R/L}$ (at the sweet spots). With finite tunneling strength and $\phi = \pi$, the qubit eigen states are $|\qubitm\rangle = \sin \frac{\eta}{2} |L\rangle + i \cos \frac{\eta}{2} |R\rangle$, $|\qubitp\rangle = \cos \frac{\eta}{2} |L\rangle - i \sin \frac{\eta}{2} |R\rangle$ with angle 
\begin{equation}
\tan \eta = \frac{2 \tau}{(U_R - U_L)/2 -(\Gamma_R - \Gamma_L)},
\label{eq:eta_SS_without_Zeeman}
\end{equation}
and energies $E_\rho = -(U_R + U_L)/4 - (\Gamma_R + \Gamma_L)/2 + {\rm s}_{\rho} \hbar \omega_0 / 2$, 
where $\rho = 0, 1$ labels the qubit states $|0, 1\rangle$, ${\rm s}_0 = -1,\, {\rm s}_1 = 1$,
and the qubit eigen frequency at the sweet spot is
\begin{align}
    \label{eq:omega0}
    \hbar \omega_0 & = \frac{2 \tau}{\sin \eta} \\ \nonumber
    & = \text{sign}(\eta) \sqrt{((U_R - U_L)/2 - (\Gamma_R - \Gamma_L))^2 + 4 \tau^2}
\end{align}
Notice that a left-right symmetric system is always in the strong tunneling regime for finite tunneling. A positive (negative) qubit frequency indicates that $|0\rangle$ ($|1\rangle$) is the qubit ground state.

{\em Qubit encoding.---}
Throughout the paper we consider a qubit encoding in terms of the qubit eigen states at the operating point. This choice allows to discuss both weak- and strong-tunneling regimes on equal footing. 

To summarize this section, setting both quantum dots to the sweet spot leads to a first-order insensitivity of the qubit frequency to dot potentials. 
The qubit states are decoupled a $\phi=0$ and $x$-axis rotations rotations are optimal at phase bias $\phi = \pi$.
Moreover, for weak tunneling ($|\eta| \ll 1$) the system is protected against relaxation, but sensitive to dephasing from fluctuations of the energy difference between the even and odd states of the individual dots. In contrast, for strong tunneling ($|\eta| \simeq \pi/2$)  the sensitivity to these fluctuations is suppressed by a factor $1/|\tan \eta| \ll 1$ 
(but the qubit is no longer protected against relaxation). Analytic results for the decoherence rates are presented in Sec.~\ref{sec:noise}.

The protection from dephasing follows from the dispersion of the qubit spectrum \cite{IthierPRB2005}. Figure~\ref{fig:1}(c) and (d) show the many-body spectra as a function of the level energy $\varepsilon_R$ for the two regimes. Analogous results hold for $\varepsilon_L$. For weak tunneling [$|\eta| \ll 1$, Figure~\ref{fig:1}(c)] the slope of the energy of the two states $|L\rangle$ and $|R\rangle$ as a function of $\varepsilon_R$ aligns at the sweet spot $\varepsilon_R = - U_R/2$. At this point, the qubit frequency ($\hbar \omega_0$, given by the energy difference of the two lowest states) is insensitive to first order in $\varepsilon_R$. For strong tunneling [$|\eta| \simeq \pi/2$, Figure~\ref{fig:1}(d)] the fluctuations of qubit frequency are further suppressed by a factor $|(U_R-U_L)/2 - (\Gamma_R - \Gamma_L)|/\tau$.

\subsection{Initialization and read-out}
\label{sec:qubit_initialization}

The operation of our qubit proposal is restricted to total odd fermion parity in the two quantum dots. Changing the total parity in the pair of quantum dots requires a quasiparticle from the superconducting leads to enter the quantum dots, similar to  a quasiparticle poisoning event \cite{GlazmanSciPostPhysLectNotes2021Jun}. It can be expected that there are (almost) no quasiparticles present in the superconducting leads for temperature much below their superconducting gap. However, recent experiments \cite{PaikPhysRevLett2011Dec, RisteNatCommun2013May, JinPhysRevLett2015Jun, SerniakPhysRevLett2018Oct, VepsalainenNature2020Aug} have shown that superconductors exhibit a density of ``hot'' quasiparticles at high energy that persists for small temperatures and dominates over thermally excited quasiparticles below $T \approx 35$mK \cite{JinPhysRevLett2015Jun} or $T \approx 150$mK \cite{PaikPhysRevLett2011Dec, RisteNatCommun2013May, SerniakPhysRevLett2018Oct}. Below, we discuss two ways to control fermion-parity changing events to initialize the fermion parity of the individual quantum dots.

{\em Initialization by detuning $\varepsilon_\nu$.---}
Depending on the system parameters, it may be energetically favorable for a quasiparticle from the environment to enter a quantum dot and, thereby, flip its fermion parity. For large charging energy $U > \Gamma_\nu$, the ground state of the quantum dot switches from odd fermion parity around the sweet spot (in our model: $(\varepsilon_\nu - U_\nu/2)^2 \leq \Gamma_\nu^2$) to even at larger level energy $\varepsilon_\nu$ \cite{BauerJPhys:CondensMatter2007Nov,MengPhysRevB2009Jun,LeeNatNanotechnol2014Jan,KirsanskasPhysRevB2015Dec}. The energy released by a quasiparticle entering the quantum dot depends on the parity of the quantum dot, resulting in different quasiparticle trapping rates \cite{OlivaresPhysRevB2014Mar}. This transition has been applied experimentally to initialize the fermion parity in a quantum dot \cite{BargerbosPRXQuantum2022Jul}. A follow-up work, Ref.~\onlinecite{PitaVidalNatPhys2023Aug}, applied this procedure to initialize a quantum dot in the odd-parity sector, with measured even-to-odd and odd-to-even switching rates of $17\ \text{kHz}$ and $0.36\ \text{kHz}$, respectively. If the fermion parity lifetimes cannot be tuned to differ significantly, one can alternatively monitor the fermion parity on the quantum dots in real time.

{\em Initialization by microwave drive.---}
Alternatively, the local fermion parity can be polarized in the odd parity state by a microwave pulse on a local gate that supplies the energy to split a Cooper pair from the condensate into one electron in the dot and one in the continuum of the superconductor \cite{WesdorparXiv2021Dec}. Similarly, the local fermion parity can be polarized in the even state by a microwave pulse that supplies the energy to excite an electron from the dot into the continuum of the superconductor. 

{\em Comment on spin initialization.---}
In the absence of Zeeman fields, the above initialization procedures do not favor a particular spin direction in the quantum dot. In this case, the spin of the quasiparticle is irrelevant and plays no role during operations. 

{\em Read-out by charge measurements.---}
The qubit state can be read out by a converting the parity information to charge. This is done by detuning the level energy of one of the quantum dots away from the sweet spot  (to $|\varepsilon_\nu + U_\nu/2| \gtrsim \Gamma_\nu$), resulting in a different charge in the dots for the even and odd parity sectors. Once tuned away, the state can be read by conventional charge-detection methods \cite{SchoelkopfScience1998, AassimePRL2001, ElzermanPRB2003, BlaisPhysRevA2004Jun, vanDrielarXiv2023Nov}. 

\subsection{Single-qubit gates}
\label{sec:qubit_rot}

\begin{figure}[t!]
\includegraphics[width=\linewidth]{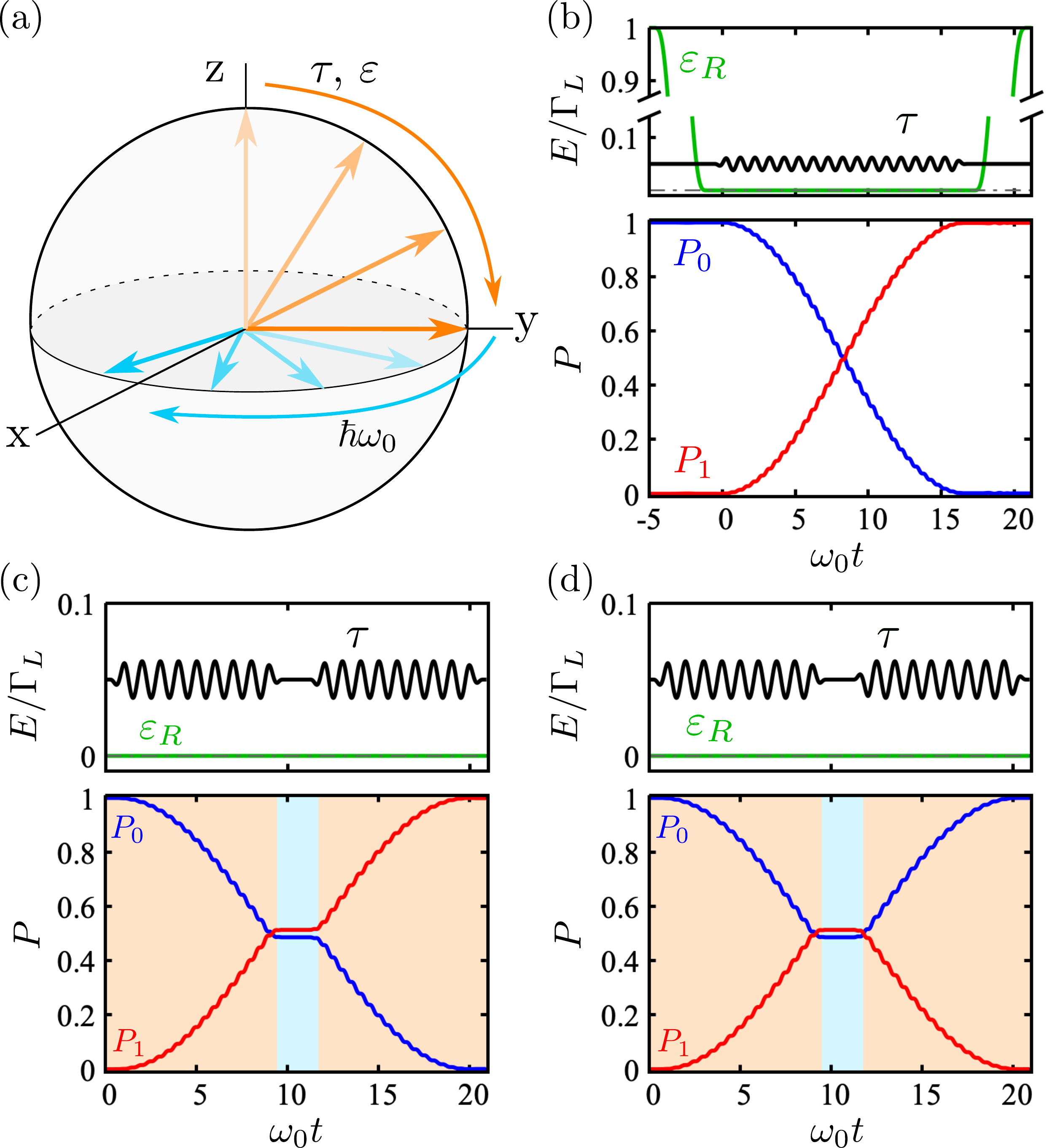}
\caption{\label{fig:2}
Two axis control of the parity qubit operated at weak tunneling $|\eta| \ll 1$. (a) Sketch of the rotations. A drive of the tunneling $\tau$ or detuning $\varepsilon_\nu$ performs $X$ or $Y$ rotations. The energy difference between the two qubit states leads to a natural rotation around $Z$ with the qubit frequency $\omega_0$. 
(b) Pauli $X$-gate by driving the tunnel coupling $\tau$ at the sweet spot. The probabilities $P_{0,1}(t) = |\langle 0,1|\Psi(t)\rangle|^2$ describe the overlap of the driven wave function $|\Psi(t)\rangle$ with the qubit eigen states $|\qubitp\rangle \approx |L\rangle$ and $|\qubitm\rangle \approx |R\rangle$ (for weak tunneling).
(c) $X_{\pi/2}Z_{2n\pi}X_{\pi/2}  = X_\pi$ sequence of rotations, the waiting time is chosen as $t_{\rm w} = n/\omega_{0},\ n=1,2,3...$ such that $Z_{2n\pi} = \textbf{1}$. 
(d) Shifting the phase of the second pulse by $\phi_0 = \pi$ rotates the qubit back to its initial configuration, $X_{\pi/2}Z_{2n\pi}X_{- \pi/2}  = \textbf{1}$.
In (b) to (d), we include a ramp in $\varepsilon_R$ at the beginning and end of the qubit operation, which simulates detuning the quantum dot away from the sweet spot for initialization and read-out. 
Hamiltonian parameters are the same as in Fig.~\ref{fig:1}(c). The qubit frequency at the sweet spot is $\hbar \omega_0 \approx 0.1414 \ \Gamma_L$. The parameterization of the pulses is contained in App.~\ref{app:pulse_parameters}. }
\end{figure}

Single-qubit gates can be achieved by driving either the tunneling amplitude between the dots $\tau(t) = \tau + \delta\tau(t)\cos\left(\Omega t + \varphi_0\right)$ or the level energy of one of the quantum dots $\varepsilon_\nu(t) = \varepsilon_{\nu} + \delta\varepsilon_{\nu}(t)\cos\left(\Omega t + \varphi_0\right)$ at the resonance frequency $\Omega =  \omega_{0} = (E_\qubitp - E_\qubitm)/\hbar$ [Fig.~\ref{fig:2}(a)]. The two computational qubit eigen states $|\qubitp\rangle$ and $|\qubitm\rangle$ are given by the two lowest-energy eigen states in the global odd-parity sector of $\hat{H}(t)$ [defined around Eq.~\eqref{eq:eta_SS_without_Zeeman}].
The amplitudes $\delta\tau, \delta\varepsilon$ determine the Rabi frequency, and the phase $\varphi_0$ sets the axis of rotation within the $X$-$Y$-plane \cite{KrantzAPR2019}. We obtain the response of the qubit states to the driving protocol from the exact unitary Schrödinger evolution with respect to $\hat{H}(t)$ $[$Eq.~\eqref{eq:full_Hamiltonian}$]$.  

{\em Weak tunneling regime, $|\eta| \ll 1$.---}
In the weak tunneling regime, the qubit eigen basis $|\qubitpm\rangle$ is approximately given by the product states $|0\rangle \approx |L\rangle$ and $|1\rangle \approx |R\rangle$, up to perturbative corrections in $|\tan \eta| \ll 1$. In this basis, the tunnel coupling $\hat{H}_T$ at $\phi = \pi$ is off-diagonal, $\hat{H}_T|L\rangle = |R\rangle$. Thus, a drive in the tunnel strength leads to qubit rotations.  

The qubit rotation by driving $\tau$ is demonstrated in Fig.~\ref{fig:2}(b). There, we furthermore include a ramp of the level energy $\varepsilon_R$ at the beginning and end of the protocol to demonstrate a read-out scenario where adiabatic detuning of $\varepsilon_R$ changes the mean charge of the even-parity state detectable by a charge measurement. The small modulation on top of the Rabi oscillations is due to the Bloch-Siegert effect \cite{BlochPhysRev1940Mar} occurring for sizable ratios $\delta\tau/\Omega$ between Rabi and driving frequency. These oscillations are often neglected when employing a rotating-wave approximation for driven quantum systems, but are included in our exact time-evolution.

Due to the finite energy difference between the two qubit states, any superposition between the qubit states Larmor precesses at frequency $\omega_{0}$ around the $Z$-axis $[$Fig.~\ref{fig:2}(a)$]$.
Changing the phase $\phi_0$ of the pulse relative to the Larmor precession of the qubit changes the axis of rotation. This is demonstrated in Figs.~\ref{fig:2}(c,d): In Fig.~\ref{fig:2}(c), two $X_{\pi/2}$ pulses are applied in sequence separated by a waiting time $t_{\rm w}$. The waiting time is chosen as $\omega_0 t_{\rm w} = 2\pi n$ such that the rotation $Z_{\omega_0 t_{\rm w}} = \textbf{1}$. The result is $X_{\pi/2} Z_{\omega_0 t_{\rm w}} X_{\pi/2} = X_{\pi}$. In Fig.~\ref{fig:2}(d), the phase of the second pulse is shifted by $\phi_0 = \pi$ such that its rotation in the opposite direction $X_{-\pi/2}$ brings the qubit back into its initial state.

\begin{figure}[t!]
\includegraphics[width=\linewidth]{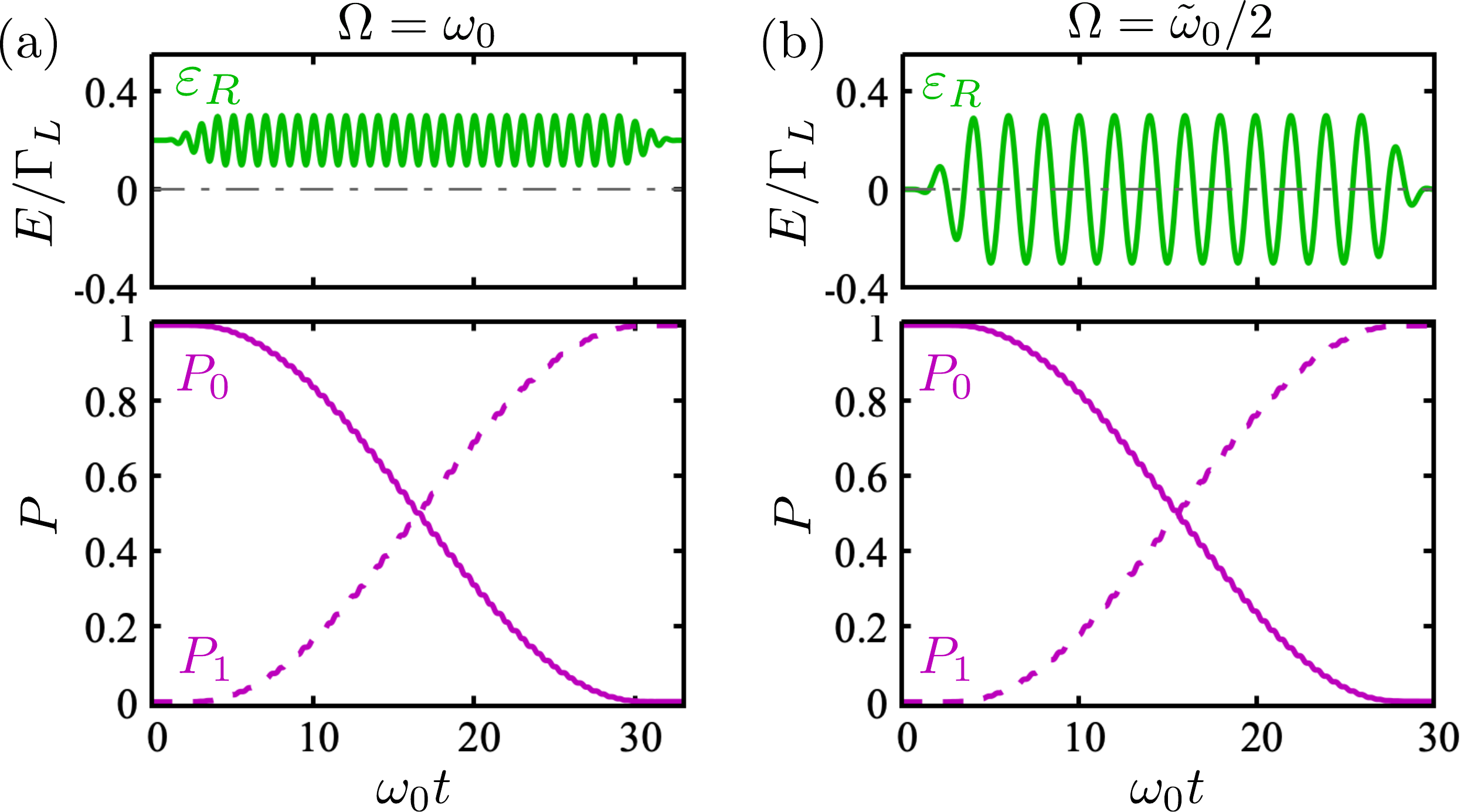}
\caption{\label{fig:3}
Pauli $X$ gate for strong tunnel coupling $|\eta| \approx \pi/2$, implemented by (a) driving $\varepsilon_R$ away from the sweet spot $\varepsilon_R = -U_R/2 + \delta \varepsilon_{R, {\rm detune}}$ in resonance $\Omega = \omega_{0}(-U_R/2 + \delta \varepsilon_{R, {\rm detune}})$, or by (b) driving $\varepsilon_R$ at the sweet spot at half of the qubit frequency averaged over one period of the drive, $\Omega = \frac{\tilde{\omega}}{2} = \frac{1}{2}(\omega_0 + \frac{1}{8}\frac{\partial^2 \omega_0}{\partial \varepsilon_\nu^2} \delta \varepsilon^2_\nu)$. 
Hamiltonian parameters are the same as in Fig.~\ref{fig:1}(d). The qubit frequency at the sweet spot is $\hbar \omega_0 \approx 0.5099 \ \Gamma_L$. The parameterization of the pulses is contained in App.~\ref{app:pulse_parameters}. 
}
\end{figure}

{\em Strong tunneling regime, $|\eta| \approx \pi/2$.---}
For strong tunneling, the qubit states $|\qubitpm\rangle$ are approximately equal superpositions of the product states $|L\rangle$ and $|R\rangle$, up to perturbative corrections in $\cot \eta$. In this case, driving the amplitude $\tau$ only affects weakly the qubit states via perturbative processes in $\cot(\eta)$. Instead, driving the level energy $\varepsilon_\nu$ of one of the quantum dots strongly couples to the qubit states. A numerical demonstration of the resulting Rabi oscillations is shown in Fig.~\ref{fig:3}.
Away from the sweet spot, the optimal driving frequency for Rabi processes equals to the qubit frequency. At the sweet spot, the optimal driving frequency to achieve complete population transfer is shifted to $\Omega = \frac{1}{2}(\omega_0 + \frac{1}{8}\frac{\partial^2 \omega_0}{\partial \varepsilon_\nu^2} \delta \varepsilon^2_\nu)$. The second term in the previous expression accounts for the shift of the mean qubit frequency at the sweet spot in the presence of the drive with amplitude $\delta \varepsilon_\nu$ in second-order perturbation theory, see App.~\ref{app:Rabi_sweet_spot} for a derivation. The derivative of the qubit frequency $\frac{\partial^2 \omega_0}{\partial \varepsilon_\nu^2}$ follows directly from the perturbative result contained in Eq.~\eqref{eq:omega01^(2)_detuning} below. Again, two-axis control is achieved by setting the phase $\phi_0$ of the pulse or, equivalently, by Larmor precession due to the energy difference of the two qubit states.

\subsection{Two-qubit gates}
\label{sec:qubit_twoqubitgates}

\begin{figure}[b!]
\includegraphics[width=1\linewidth]{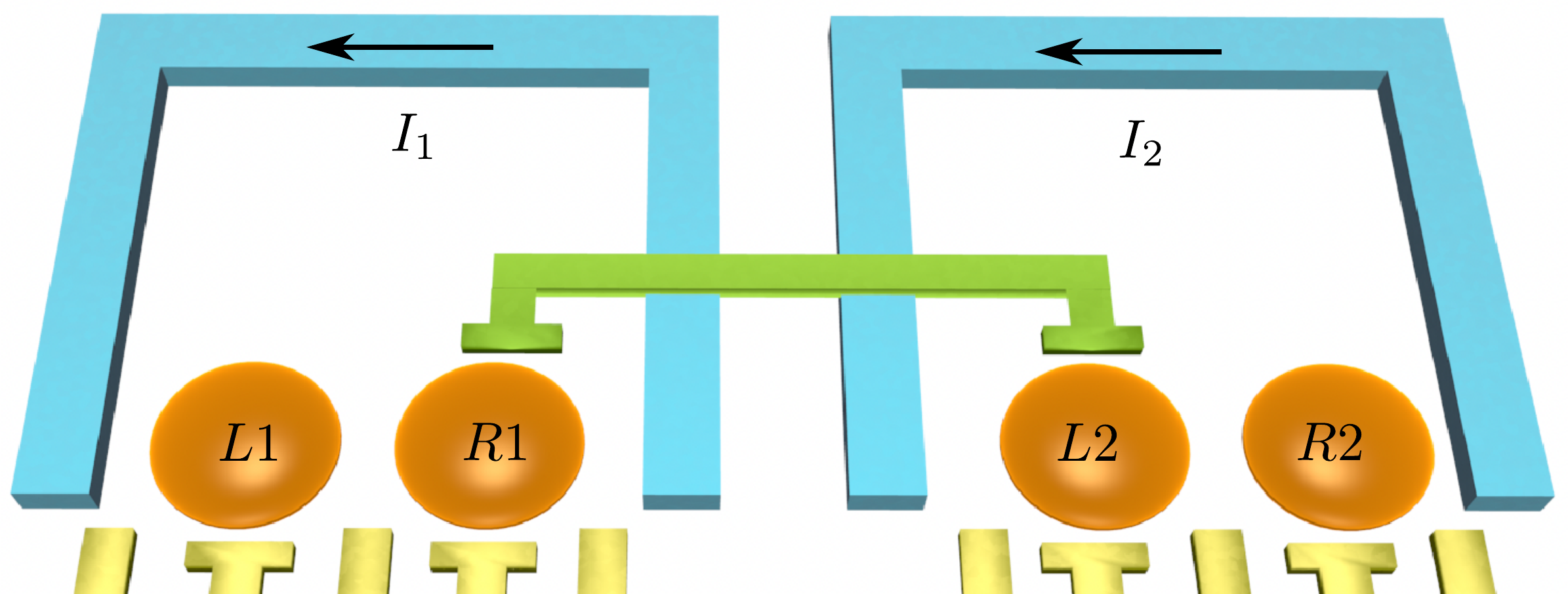}
\caption{\label{fig:5} Device sketch for two-qubit gates: Two fermion-parity qubits $1, 2$ can be coupled inductively or by a floating gate (green) mediating capacitive between quantum dots of distinct qubits. 
}
\end{figure}

We describe two-qubit gates that arise from inductive coupling of the superconducting loops or capacitive coupling between quantum dots of two distinct qubits indexed by $j = 1,2$. Figure~\ref{fig:5} shows a sketch for a setup allowing to realize two-qubit gates.

{\em Capacitive coupling.---}
Mutual capacitive coupling between quantum dots $\nu_1$, $\nu_2$ of adjacent qubits can be described by an interaction term $U_{12} \hat{n}_{\nu_1,1} \hat{n}_{\nu_2,2}$, where $\hat{n}_{\nu_j, j} = \sum_\sigma \hat{n}_{\sigma, \nu_j, j}$ is the occupation of the $\nu_j = \text{L, R}$ quantum dot of qubit $j = 1,2$. For concreteness, we consider capactive coupling between the right quantum dot $\nu_{R,1}$ of qubit $1$ and the left quantum dot $\nu_{L,2}$ of qubit $2$. Using $\hat{n}_{\nu,j} = \frac{d \hat{H}_j}{d \varepsilon_{\nu, j}}$ with $\hat{H}_j$ being the Hamiltonian Eq.~\eqref{eq:full_Hamiltonian} for each qubit, we apply the Hellman-Feynman theorem $\langle n | \frac{d\hat{H}}{d\varepsilon_{\nu, j}}|n\rangle = \frac{d E_n}{d\varepsilon_{\nu, j}}$ where $|n\rangle$ labels the $n$'th eigen state. The capacitive coupling projected onto the qubit eigenspace is
\begin{equation}
{\cal P}\, U_{12}\hat{n}_{R1}\hat{n}_{L2}{\cal P}=\hbar^2 U_{12} \frac{d \omega_{0,1}}{d \varepsilon_{R,1}}\frac{d \omega_{0,2}}{d \varepsilon_{L,2}} \hat{\rho}_{1}^{z}\otimes\hat{\rho}_{2}^{z},
\label{eq:two_qubit_capacitively}
\end{equation}
where $\hat{\rho}_{j}^z$ are Pauli-$z$ operators in the space of eigen states $|\rho_j \rangle_j,\ \rho_j = \qubitpm$ of qubit $j$ (see around Eq.~\eqref{eq:eta_SS_without_Zeeman} for a definition of the qubit eigen states at the operating point), ${\cal P} = \sum_{{\rho}_1, {\rho}_2} |{\rho}_1\rangle_1 |{\rho}_2\rangle_2 \langle {\rho}_1|_1 \langle {\rho}_2 |_2$ the projector onto these states, and the qubit frequency $\hbar \omega_{0,j}$ to second order is given in Eq.~\eqref{eq:omega01^(2)_detuning} below. At the sweet spot, $\varepsilon_{\nu, j} = -U_{\nu, j}/2$, the capacitive coupling does not differentiate between the qubit states as the charge dipole moment of both qubit states vanishes. The charge dipole moment increases linearly with the detuning of the level energy $\varepsilon_{\nu, j}$ away from the sweet spot, which allows to switch on the two-qubit coupling using electrostatic control of the level energy $\varepsilon_{\nu, j}$. \footnote{We note that single-qubit gates in the strong tunneling regime could also be performed by detuning $\varepsilon_\nu$. If the detuned quantum dot is capacitively coupled to a quantum dot from another qubit, a two-qubit rotation results only if the other quantum dot is also detuned (see Eq.~\eqref{eq:two_qubit_capacitively}). To avoid accidental two-qubit gates when performing single-qubit operations, one can consider a design with divided tasks where only one quantum dot of each qubit is used for read-out and single-qubit gates by detuning $\varepsilon_\nu$, while the other quantum dot is capacitively coupled to other qubits.} Long distance capacitive coupling between quantum dots can be mediated by floating gates \cite{ChanApplPhysLett2002, ChanPhysicaE2003}, schematically depicted in Fig. \ref{fig:5} in green.

{\em Inductive coupling.---}
Inductive coupling via the superconducting loops is described by a term $\hat{H}_L = L\, \hat{I}_1 \hat{I}_2$ where $\hat{I}_j = \frac{2e}{\hbar}\frac{d\hat{H}_j}{d\phi_j}$ is the supercurrent operator in the loop and $L$ the mutual inductance. However, the inductive coupling is not well-suited for our qubit design because at the phase sweet spot $\phi = \pi$, the derivatives, and thereby the inductive coupling, are zero. Moreover, around $\phi = \pi$, the coupling is proportional to $\tau \sin \eta$ [see Eq.~\eqref{eq:omega01^(2)_phi}] and, thus, suppressed in the weak tunneling regime $|\eta| \ll 1$. It therefore requires strong tunneling between the dots to give a significant contribution. 

\section{Parameter noise}
\label{sec:noise}

In this section, we quantify the susceptibility of the parity qubit to noise around the sweet spot. We use Fermi's golden rule and the Bloch-Redfield approximation to determine relaxation and dephasing rates. We use lowest-order perturbation theory for the qubit frequency $\omega_0$ as a function of the fluctuating parameters. Here we focus on what we believe are the most relevant noise parameters. A discussion including fluctuations in all system parameters is contained in App.~\ref{app:decoherence}.

We include terms coupling to the local spin,
\begin{equation}
\hat{H}^B_\nu = \sum_\sigma {\rm s}_\sigma B_{z, \nu} \hat{n}_{\sigma \nu} +  \left[(B_{x,\nu}-iB_{y,\nu})\hat{c}_{\uparrow \nu}^{\dagger}\hat{c}_{\downarrow \nu}+\text{H.c.}\right],
\end{equation}
with ${\rm s}_\sigma = \pm 1$ for $\sigma = \uparrow,\downarrow$, 
which describes Zeeman coupling to magnetic fields as well as spin-spin exchange coupling to a nuclear spin bath.
We further include spin-orbit coupling, which replaces the tunneling term in Eq.~\eqref{eq:H_tau},
\begin{equation}
\hat{H}_{T}^\text{SOC}= \tau e^{i \phi}  \left[\hat{c}_{\uparrow L}^{\dagger},\ \hat{c}_{\downarrow L}^{\dagger}\right]e^{i \theta \vec{n} \cdot \vec{\sigma}}\left(\begin{array}{c}
\hat{c}_{\uparrow R}\\
\hat{c}_{\downarrow R}
\end{array}\right)+\text{H.c.}
\label{eq:H_T^SOC}
\end{equation}
with the vector of Pauli matrices $\vec{\sigma} = (\sigma_x, \sigma_y, \sigma_z)^{\rm T}$ in spin space.
The matrix $e^{i \theta \vec{n} \cdot \vec{\sigma}}$ describes a rotation of the electron's spin by an angle $\theta$ around the axis $\vec{n}$ as they tunnel from the right to left dot. For $\theta = 0$, the spin-orbit coupling is zero and Eq.~\eqref{eq:H_tau} is recovered. The axis $\vec{n}$ is called the spin-orbit direction. By choosing the spin quantization axis in both quantum dots to be aligned with $\vec{n}$, the spin-orbit coupling becomes diagonal, $\tau e^{i \theta \sigma_z}$. We choose this basis for the following, keeping in mind that the Zeeman fields are now given with respect to this basis, {\it i.e.} $B_z$ points parallel to the spin-orbit axis $\vec{n}$.
In a device implementation, we expect the direction of the externally applied field to be constrained by the superconductor geometry and $g$-factor anisotropy of the semiconductor, see the discussion in Sec.~\ref{sec:gate_times} for details.

In the presence of Zeeman fields, the angle $\eta$ determining the qubit eigen states and operating regimes (as defined in Eq.~\eqref{eq:eta_SS_without_Zeeman}) is modified,
\begin{equation}
\tan \eta_{\sigma,\lambda} = \frac{2 \tau}{(U_R - U_L)/2 + {\rm s}_\sigma(B_{z L} - B_{z R}) + {\rm s}_\lambda(\Gamma_R - \Gamma_L)}
\label{eq:eta_SS_basis_Hnu}
\end{equation}
where ${\rm s}_\lambda = \pm 1$ for $\lambda = \lambdap, \lambdam$. 
The qubit eigen frequency is accordingly
\begin{equation}
\hbar \omega_{0} = 2\frac{\tau}{\sin{\eta_{\sigma,\lambdam}}}\, .
\label{eq:omega0_with_Zeeman}
\end{equation} 
Equations~\eqref{eq:eta_SS_without_Zeeman} and \eqref{eq:omega0} are recovered from Eqs.~\eqref{eq:eta_SS_basis_Hnu} and \eqref{eq:omega0_with_Zeeman} for $B_{z,L}=B_{z,R}$ and $\lambda=\lambdam$, see Sec.~\ref{sec:qubit_regimes} \footnote{This equation is valid in the presence of an exact or approximate $U(1)$ spin-rotation symmetry around the axis of an applied Zeeman field or around the spin-orbit direction, such that perpendicular components can be treated perturbatively.}.
The spin-orbit coupling angle $\theta$ does not enter in Eq.~\eqref{eq:eta_SS_basis_Hnu}. It only enters as a relative phase factor between $|L\rangle$ and $|R\rangle$ in the qubit eigen states, see App.~\ref{app:diagonalziation}.

{\em Dephasing and relaxation rates.---} Assuming the computational qubit subspace to be decoupled from the remaining states governed by $\hat{H}(t)$ at any time $t$, the dephasing rate $\Gamma_\varphi^\chi$ due to a noisy, linearly coupled parameter $\chi$ is given in Bloch-Redfield theory as \cite{IthierPRB2005, KrantzAPR2019}
\begin{equation}
\Gamma_\varphi^\chi = \pi \left( \frac{\partial \omega_{0}}{\partial \chi} \right)^2 S_\chi(\omega \to 0).
\label{eq:def_dephasing_rate}
\end{equation}
This presupposes that the noise spectral density 
\begin{equation}
S_\chi (\omega) = \int_{-\infty}^\infty d\tau \langle \chi(0) \chi(\tau) \rangle e^{- i \omega \tau}
\end{equation}
is regular near $\omega \approx 0$ up to frequencies of order of $\Gamma_\varphi$, where $\langle \chi(0) \chi(\tau) \rangle$ is the autocorrelation function of the fluctuating parameter $\chi$ with respect to its underlying statistical distribution.

The relaxation rate $\Gamma_{\text{rel}}^\chi$ is given in Fermi's golden rule \cite{IthierPRB2005, KrantzAPR2019},
\begin{equation}
\Gamma_{\text{rel}}^\chi(\omega_0) = \frac{\pi}{2 \hbar^2} \left| \langle \qubitp | \frac{d \hat{H}}{d \chi} | \qubitm \rangle\right|^2 S_\chi(\omega_{0}).
\label{eq:def_relaxation_rate}
\end{equation}
Similarly, the excitation rate is given by $\Gamma_{\text{exc}}^\chi(\omega_0)=\Gamma_{\text{rel}}^\chi(-\omega_0)$. Both processes contribute to the relaxation rate $\Gamma_1^\chi (\omega_0)= \Gamma_{\text{rel}}^\chi(\omega_0) + \Gamma_{\text{exc}}^\chi(\omega_0)$. At temperatures $k_{\rm B} T \ll \hbar \omega_{0}$, the excitation rate $\Gamma_{\text{exc}}^\chi(\omega_0)$ is exponentially suppressed~\cite{IthierPRB2005}.
Both classical and quantum fluctuations are included in this formulation via the noise power spectral density $S_\lambda(\omega)$ \cite{KrantzAPR2019}.

At first-order insensitive sweet spots, the dephasing due to noise in the respective parameter $\lambda$ is determined by higher-order terms in $\lambda$. 
Explicit expressions are provided in Ref.~\cite{IthierPRB2005}.
These contributions depend on the detailed form of both the higher-order terms and the noise power, and we do not include them here.

{\em Level energy fluctuations.---} 
Electric field fluctuations can couple to the dot-level energies $\varepsilon_\nu$. To quantify their effect, we calculate the qubit frequency, $\hbar \omega_{0}^{(2)} = E_{\qubitp}^{(2)}-E_{\qubitm}^{(2)}$, to second order in  the detuning of the level energy $\varepsilon_\nu$ away from the sweet spot,
\begin{align}
\label{eq:omega01^(2)_detuning}
\hbar \omega_{0}^{(2)}&=\hbar \omega_{0}^{(0)}
+\left(\frac{\left(\varepsilon_{L}+\frac{U_{L}}{2}\right)^2}{2\Gamma_{L}} - \frac{\left(\varepsilon_{R}+\frac{U_{R}}{2}\right)^2}{2\Gamma_{R}}\right) \cos\eta_{\sigma,\lambdam} \nonumber \\
&\ -\left(\frac{\varepsilon_{L}+\frac{U_{L}}{2}}{2\Gamma_{L}} - \frac{\varepsilon_{R}+\frac{U_{R}}{2}}{2\Gamma_{R}} \right)^{2}\tau\sin\eta_{\sigma,\lambdam}\,. 
\end{align}
where $\hbar \omega_{0}^{(0)}$ is the qubit frequency at the operating point as given by Eq.~\eqref{eq:omega0_with_Zeeman}.
The first correction arises from the quadratic dependence of the energy of the local even parity states of the individual dots on the level energy $\varepsilon_\nu + U_\nu/2$ around the sweet spot. The second correction describes a modification of the tunneling between these states due to the change of the Andreev bound state superposition by detuning $\varepsilon_\nu$. In Eq. \eqref{eq:omega01^(2)_detuning}, we neglected a term describing second-order processes involving high-energy states $|\lambdap\rangle_\nu$ with a symmetric superposition of $|0\rangle_\nu$ and $|2\rangle_\nu$, see App.~\ref{app:decoherence} for the full result. These states are separated from the qubit subspace by an energy difference of the order of $2\Gamma_\nu$.
Notice that for both weak- and strong-tunneling regimes, there is a common-mode fluctuation given by the zeros of the factors of the two correction terms in Eq.~\eqref{eq:omega01^(2)_detuning} that leaves the qubit frequency invariant, and thus does not lead to dephasing.

The matrix element $\langle \qubitp | \frac{d\hat{H}}{d\varepsilon_\nu} | \qubitm \rangle = 0$ at the sweet spot. 
Thus we have $\Gamma_{\text{rel}}^{\varepsilon_\nu} = 0$
independent of the tunneling strength.

{\em Tunneling strength $\tau$.---} Variations in the magnitude of the tunneling amplitude $\tau \to \tau + \delta \tau$ enter in the qubit frequency up to second order as
\begin{align}
\label{eq:omega01^(2)_tau}
\hbar \omega_{0}^{(2)}&=\hbar \omega_{0}^{(0)} + 2 \delta \tau \sin \eta_{\sigma,\lambdam} \\ \nonumber
 & \, + 2  (\hbar \omega_{0}^{(0)})^3 \left( (U_R - U_L)/2 - (\Gamma_R - \Gamma_L) \right)^2 (\delta \tau)^2.
\end{align}
The resulting dephasing and relaxation rates are 
$\Gamma_{\varphi}^{\tau} = \frac{4 \pi}{\hbar^{2} }\sin^{2}\eta_{\sigma,\lambdam} S_{\tau}(\omega\to0)$ and
$\Gamma_{\text{rel}}^{\tau}  = 2 \frac{\pi}{\hbar^{2}}\cos^{2}\eta_{\sigma,\lambdam}S_{\tau}(\omega_{0})$.
For strong tunneling, $|\eta| \approx \pi/2$, the term linear in $\delta \tau$ dominates dephasing.
For weak tunneling, $|\eta| \ll 1$, the linear term is suppressed and the equation simplifies,
\begin{equation}
    \hbar \omega_{0}^{(2)} \approx \hbar \omega_{0}^{(0)} + \frac{2 (\delta \tau)^2}{ (U_R - U_L)/2 - (\Gamma_R - \Gamma_L)} \, .
\end{equation}
Thus, for weak tunneling $|\eta| \ll 1$, the system has a sweet spot protecting against dephasing from electric field fluctuations coupling to $\tau$.

{\em Phase fluctuations.---} The phase difference between the superconductors can fluctuate due to magnetic flux variations. To second order in the superconducting phase difference, the qubit frequency can be written as
\begin{equation}
\hbar \omega_{0}^{(2)}=\hbar \omega_{0}^{(0)} 
-\frac{\delta\phi^2}{4}\tau\sin\eta_{\sigma,\lambdam}\,.
\label{eq:omega01^(2)_phi}
\end{equation}
Also here, we neglected a term describing second-order processes via the high-energy states $|\lambdap\rangle_\nu$, {\em c.f.} App.~\ref{app:decoherence}.
At the sweet spot $\varepsilon_\nu + U_\nu/2 = 0$ and $\phi = \pi$, also the matrix element $\langle \qubitp | \frac{d \hat{H}}{d \phi} | \qubitm \rangle = 0$, such that both Bloch-Redfield dephasing and the relaxation rate in Fermi's golden rule are zero. 

{\em Magnetic field fluctuations.---} Fluctuating magnetic fields coupling to the electron spin include Zeeman coupling to external magnetic fields as well as spin-spin exchange interaction with the nuclear spin bath. Magnetic fields along the spin-orbit axis have a linear contribution to the qubit frequency
\begin{equation}
\hbar \omega_{0}^{(1)}=\hbar \omega_{0}^{(0)} +{\rm s}_\sigma (B_{z, L} - B_{z, R}) \cos\eta_{\sigma,\lambdam}.
\label{eq:omega01^(2)_Bz}
\end{equation}
This linear contribution is large for weak tunneling $|\eta_{\sigma,\lambdam}| \ll 1$, but goes to zero for strong tunnel coupling with $|\cot \eta_{\sigma,\lambdam}| \ll 1$.
Common-mode fluctuations $B_{z, L} = B_{z, R}$ do not contribute to dephasing.
The relaxation rate $\Gamma_{\text{rel}}^{B_{z,\nu}} =\frac{\pi}{8\hbar^{2}}\sin^{2}\eta_{\sigma,\lambdam}S_{B_{z,\nu}}(\omega_{0})$ behaves oppositely: It is large for strong tunneling, but approaches zero for weak tunneling with $|\tan \eta_{\sigma,\lambdam}| \ll 1$.

Magnetic field fluctuations perpendicular to the spin-orbit direction induce transitions between the two spin sectors. Choosing one of the spin sectors as the computational space, these fluctuations would take the qubit out of the computational space. This can be avoided by applying an external Zeeman field $B_z^{\text{ext}}$. Ideally, this Zeeman field should be aligned with the spin-orbit direction $\vec{n}$ to avoid spin-flipping processes from the spin-orbit coupling. Then, field fluctuations in the orthogonal directions $\delta B_\perp$ enter only to second order,
\begin{equation}
\hbar \omega_{0}^{(2)}=\hbar \omega_{0}^{(0)} - \frac{\delta B_{\perp}^{2}}{(B_{z}^{\text{ext}})^{2}}\frac{\tau\sin(\eta_{\sigma,\lambdam})\sin^{2}\theta}{1-\tau^{2}(\sin\eta_{\sigma,\lambdam})^{-2}(B_{z}^{\text{ext}})^{-2}}\,.
\label{eq:omega01^(2)_Bx}
\end{equation}
This dephasing contribution decreases quadratically with the applied magnetic field and requires spin-orbit coupling, as expressed by the proportionality $\sin^2(\theta)$ to the spin-orbit angle $\theta$. Also here, the perpendicular fluctuations do not contribute to the relaxation rate in Fermi's golden rule, $\Gamma_{\text{rel}}^{B_\perp} = 0$.

{\em Remaining parameters.---} While the parity qubit is linearly protected against fluctuations in the level energies and the magnetic flux, fluctuations in other parameters can affect qubit performance. Fluctuations of local parameters of one of the dots, such as the charging energy $U_\nu$ and the induced pairing $\Gamma_\nu$, contribute to the dephasing rate with a term proportional to $\sin^2(\eta_{\sigma,\lambdam})$ and to the relaxation rate proportional to $\cos^2(\eta_{\sigma,\lambdam})$. 

Furthermore, a mutual charging energy $U_\text{LR} \sum_{\sigma, \sigma'} \hat{n}_{\sigma L} \hat{n}_{\sigma' R}$ modifies the system Hamiltonian in the odd fermion-parity subspace by replacing $U_\nu \to U_\nu + 2 U_\text{LR}$. Accounting for the modified sweet spot, fluctuations in $U_\text{LR}$ enter only to second order in the qubit frequency by detuning the system from the sweet spot as described by Eq.~\eqref{eq:omega01^(2)_detuning}. 

Beyond the decoherence channels discussed here, the relaxation rate of our proposed qubit is lower bounded by the fermion-parity lifetime of the individual quantum dots.
Previous experiments determined a fermion-parity lifetime of $T_{\rm parity} = 200\ \mu$s \cite{Janvier_Science2015}, $160\ \mu$s \cite{Hays_PRL2018}, and $22\ \mu$s \cite{Hays_Science2021}. Ref.~\cite{BargerbosPRXQuantum2022Jul} found that the parity lifetime depends sensitively on the level energy $\varepsilon$, ranging from $60\ \mu$s when the ground state of the quantum dot is a singlet to $2.8\ $ms when ground state is a doublet (see also our discussion in Sec.~\ref{sec:qubit_initialization}).

Other sources of dephasing include electron-phonon coupling which via spin-orbit coupling lead to dephasing \cite{GolovachPhysRevLett2004Jun}. This mechanism can be suppressed by applying a polarizing magnetic field, see Sec.~\ref{sec:gate_times} for the required magnetic fields and compatibility with superconducting leads. 
We leave detailed discussions of other sources of dephasing, such as electron-phonon coupling and quasiparticle effects, to further work.

\section{Qubit operation for large $U \gg \Delta$}
\label{sec:largeU}

Several material candidates for implementation of the fermion-parity qubit, such as InAs or SiGe semiconductors with Al superconductors (see also Sec.~\ref{sec:gate_times} and \ref{sec:coherence_times}), typically realize quantum dots with charging energy $U$ much larger than the superconducting gap $\Delta$.
Our proposal can also be implemented in this limit. 
We here consider the case that the hopping rate $\Gamma$ between dot and superconductor is smaller than or similar to $\Delta$. The latter requirement ensures that the Kondo temperature
$
    k_B T_K \approx 0.182 \, U \sqrt{\frac{8\Gamma}{\pi U}} e^{-\pi U/8\Gamma} 
$
\cite{BauerJPhys:CondensMatter2007Nov} is much smaller than $\Delta$, where the numerical factor $0.182$ is obtained from numerical renormalization group calculations of the Anderson model \cite{HaldaneJPhysC:SolidStatePhys1978Dec,YoshiokaJPhysSocJpn2000Jun}. In this case, the superconducting gap $\Delta$ suppresses Kondo correlations. Instead, the even-parity states in the combined dot-superconductor systems are Yu-Shiba-Rusinov singlets composed of one electron on the quantum dot and one quasiparticle from the superconductor.
Above $k_B T_K > 0.3 \, \Delta$ \cite{SatoriJPhysSocJpn1992Sep,YoshiokaJPhysSocJpn2000Jun,BauerJPhys:CondensMatter2007Nov,Grove-RasmussenNatCommun2018Jun}, 
the Yu-Shiba-Rusinov singlet crosses over to a Kondo singlet. 
We focus on the limit $\Gamma \lessapprox \Delta \ll U$.

Conceptually, operations in the Yu-Shiba-Rusinov limit can be performed as described in Sec. \ref{sec:qubit} for the infinite-gap limit. The main difference is that the even parity bound states are Yu-Shiba-Rusinov singlets instead of even-parity superpositions of zero and double occupation ({\it c.f.} Sec.~\ref{sec:qubit_regimes}). This physics can be accurately captured by taking into account only a few levels in the superconducting leads \cite{BaranPhysRevB2023Dec}.
	
For concreteness, we here consider material parameters for InAs/Al heterostructures where gate-defined quantum dots coupled to superconductors are routinely realized \cite{EstradaSaldanaPhysRevLett2018Dec,Grove-RasmussenNatCommun2018Jun,BoumanPhysRevB2020Dec,EstradaSaldanaPhysRevB2020Nov,TomDvirNature2023Feb}. We take typical values for charging energy, $U = 1.5\ $meV, and tunneling rate, $\Gamma = 0.1\ $meV, \cite{EstradaSaldanaPhysRevLett2018Dec}. Then, the Kondo temperature $k_B T_K \approx 0.3\ \mu$eV is much smaller than $\Delta$.

Here, we solve the interacting problem by reducing the superconducting lead to a single site. This reduction is known as the zero-bandwidth approximation and is justified for $k_B T_K \ll \Delta$ \cite{Grove-RasmussenNatCommun2018Jun}. We expect this model to qualitatively capture the relevant physics for $t < \Delta$, while quantitative corrections may be found when taking into account additional superconducting levels \cite{BaranPhysRevB2023Dec}. 
In this limit, the Hamiltonian for each quantum dot, including a single superconducting site created by $\hat{d}_{\sigma\nu}^\dagger$, reads
\begin{align}
\label{eq:H_nu_ZBA}
\hat{H}_{\nu}^{\rm ZBA} & = 
\sum_{\sigma}\varepsilon_{\nu}\hat{n}_{\sigma\nu} 
 + U_\nu \hat{n}_{\uparrow\nu}\hat{n}_{\downarrow\nu} \\ \nonumber 
& \, + \sum_\sigma  t_{\nu} (\hat{c}^\dagger_{\sigma \nu} \hat{d}_{\sigma \nu} + {\rm H.c.}) + \Delta \left( \hat{d}_{\uparrow\nu}\hat{d}_{\downarrow\nu}+{\rm H.c.}\right)\,,
\end{align}
where $t_{\nu}$ models the tunneling amplitude to the single-site superconductor.
The tunneling amplitude relates to the tunneling rate $\Gamma_\nu$ as defined below Eq.~\eqref{eq:H_nu} as $t_\nu \approx \sqrt{0.9618 \ \Gamma_\nu \Delta }$, where the numerical factor $0.9618$ is obtained by fitting the Nambu tunneling self energy to numerical renormalization group calculations \cite{BaranPhysRevB2023Dec}.
The full Hamiltonian is constructed as $\hat{H}^{\rm ZBA} = \hat{H}_{\nu}^{\rm ZBA} + \hat{H}_\tau$ with $\hat{H}_\tau$ given in Eq.~\eqref{eq:H_tau}. 
We take parameters $U_L = U_R = 15 t_L$, $\Delta = 2 t_L$, and $t_R = 1.1 t_L$ with $t_L = 0.1\ $meV as default. 

\begin{figure}
    \centering
    \includegraphics[width=0.7\columnwidth]{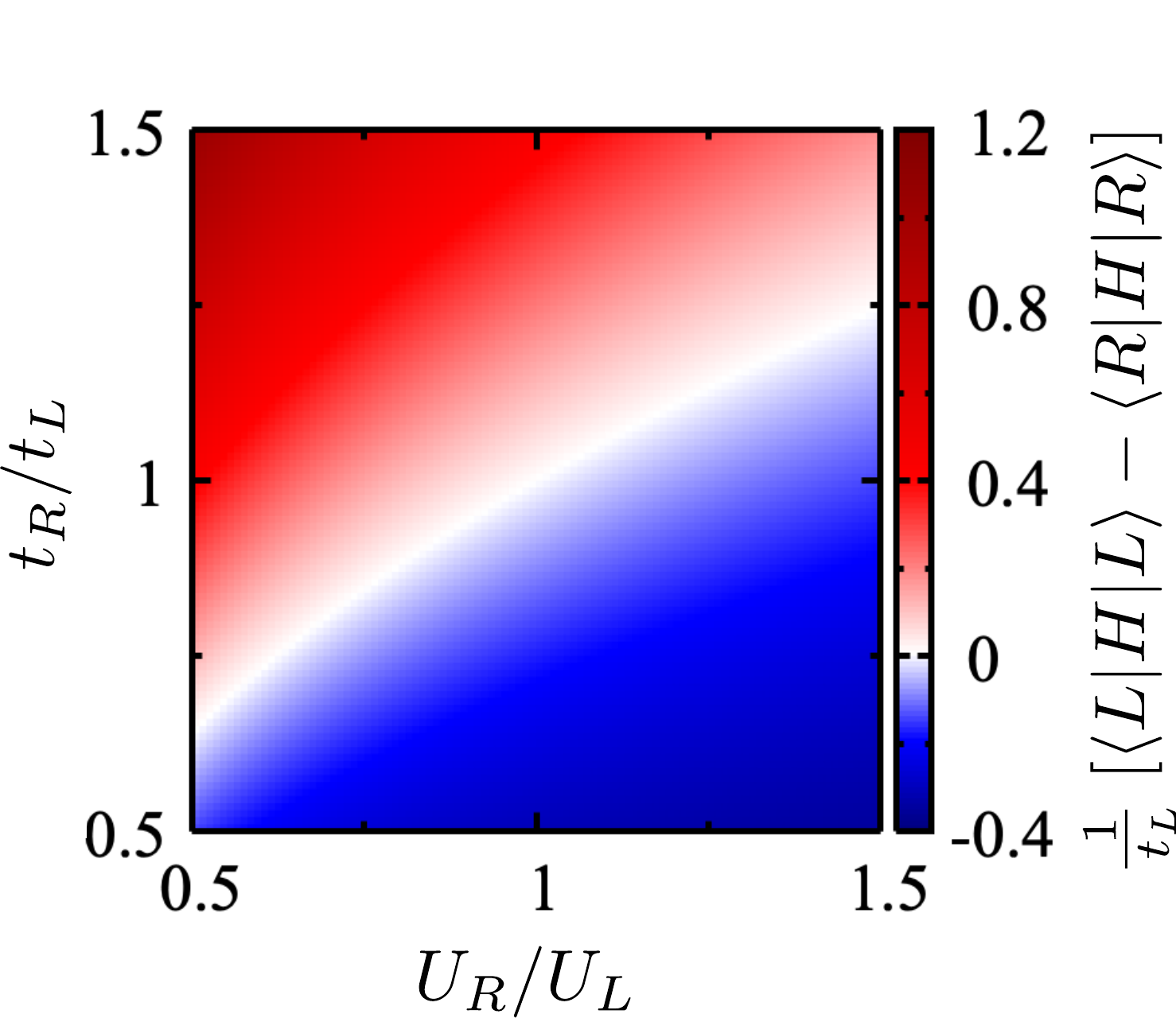}
    \caption{Energy difference of the decoupled states $\langle R | \sum_\nu H_\nu^{\rm ZBA} | R \rangle - \langle L | \sum_\nu H_\nu^{\rm ZBA} | L \rangle$ in zero-bandwidth approximation, Eq.~\eqref{eq:H_nu_ZBA}, as a function of the ratio of dot-lead tunneling strengths $t_R / t_L$ and the ratio of charging energies $U_R/U_L$ around $U_L = 15 t_L$ and at the sweet spot $\varepsilon_\nu = - U_\nu/2$.}
    \label{fig:ZBA_LR}
\end{figure}

As before, the weak- and strong tunneling regimes are distinguished by the relative strength of the interdot tunneling $\tau$ to the energy difference of the states $|L\rangle$ and $|R\rangle$. Here, the state $|L\rangle$ ($|R\rangle$) consists of the lowest-energy doublet, {\it i.e.} odd-fermion parity, state on the left (right) dot and a Yu-Shiba-Rusinov singlet on the right (left) dot. 
Figure~\ref{fig:ZBA_LR} shows the energy difference of these two states 
$E_R^{\rm ZBA} - E_L^{\rm ZBA}$ with $E_\nu^{\rm ZBA} = \langle \nu | H^{\rm ZBA} | \nu \rangle$ as a function of the asymmetry $t_R/t_L$ and $U_R/U_L$ of the two quantum dots at the sweet spot $\varepsilon_\nu = - U_\nu/2$ for $U_L = 15 t_L$. 
Notably, asymmetries in the charging energy enter the spectrum approximately proportional to $U_R/U_L$, which is much weaker than in the infinite-gap limit (c.f. Eq.~\eqref{eq:omega0}).
For devices with identically shaped quantum dots we expect $U_R/U_L \approx 1$.
We expect the asymmetry in tunneling amplitudes $|t_R/t_L - 1|$ to be generically larger than $|U_R/U_L - 1|$ and further the tunneling amplitudes $t_\nu$ may be gate-tunable. 
This motivates our choice of using an asymmetry in $t_\nu$ to model the asymmetry while taking $U_L = U_R$.
In the following, we set $\tau = 0.05 t_L < |t_L - t_R|$ for the weak tunneling regime and $|t_L - t_R| < \tau = 0.5 t_L < t_L, t_R$ for the strong tunneling regime.

\begin{figure}
    \centering
    \includegraphics[width=\columnwidth]{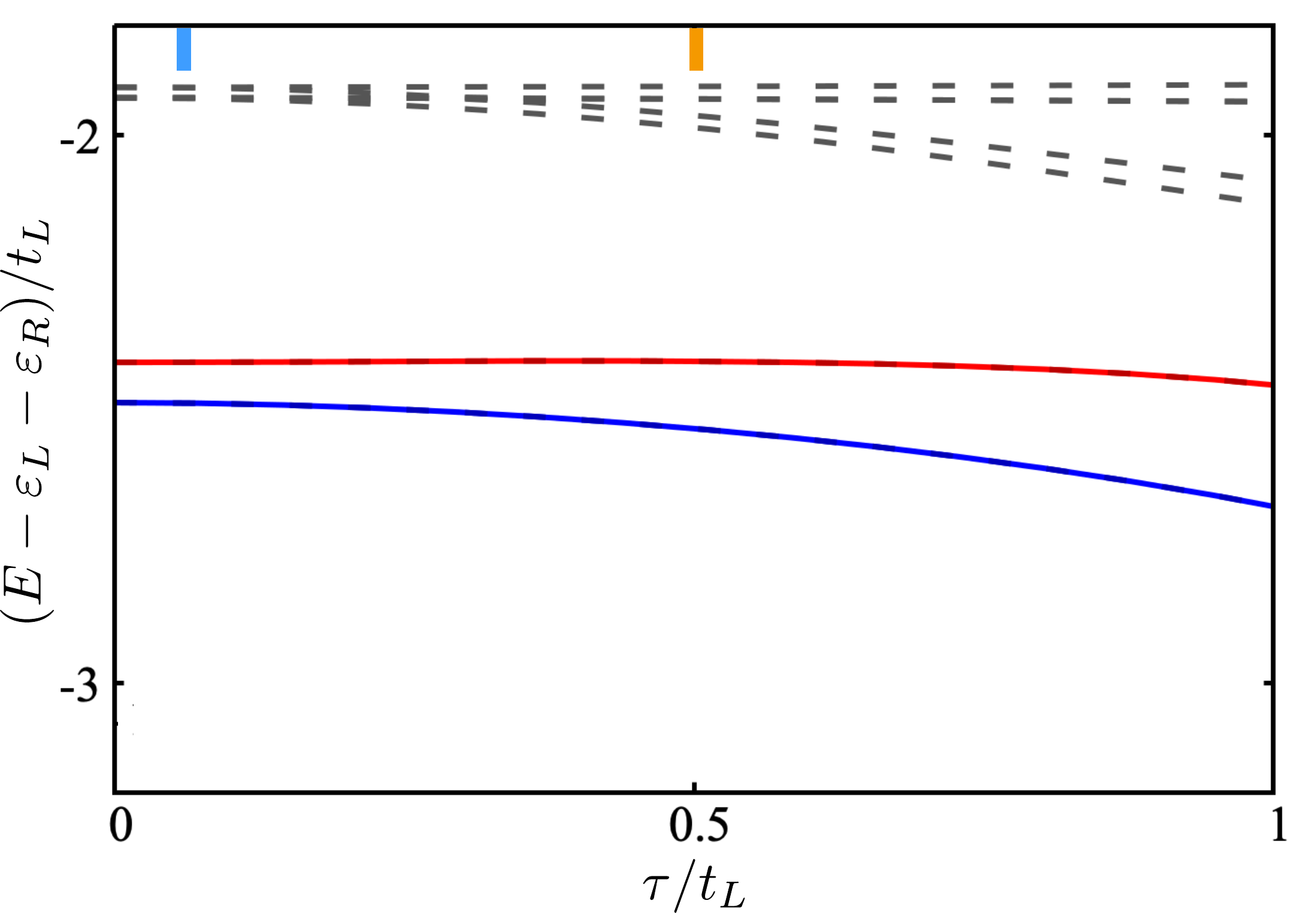}
    \caption{Spectrum as a function of level energy $\tau$ for $U_\nu \gg \Delta$ calculated in zero-bandwidth approximation at the sweet spot $\varepsilon_\nu = - U_\nu / 2$.
    The remaining parameters are $U_L = U_R = 15 t_L$, $t_R = 1.1 t_L$ and $\Delta = 2 t_L$, $\phi = \pi$. 
    The blue and red line denote the qubit ground and excited state, respectively. The gray dashed lines indicate excited states. The light blue and orange markers indicate the values $\tau = 0.05 t_L$ and $\tau = 0.5 t_L$ that we use for the weak and strong tunneling regime, respectively.}
    \label{fig:ZBA_tau}
\end{figure}

{\em Interdot tunneling strength $\tau$.---} 
The spectrum of the double-quantum dot as a function of tunneling strength is shown in Fig.~\ref{fig:ZBA_tau}. In the weak tunneling regime, the spectrum is insenstive towards fluctuations in $\tau$. For strong tunneling, the system has a linear sensitivity towards changes in $\tau$. Notably, the spectral gap at $\tau = 0.5 t_L$ is smaller than $\tau$ due to an effective suppression of tunneling between the states $|L\rangle$ and $|R\rangle$ in the Yu-Shiba-Rusinov regime due to bound-state formation with quasiparticles from the superconductor. 

\begin{figure}
    \centering
    \includegraphics[width=\columnwidth]{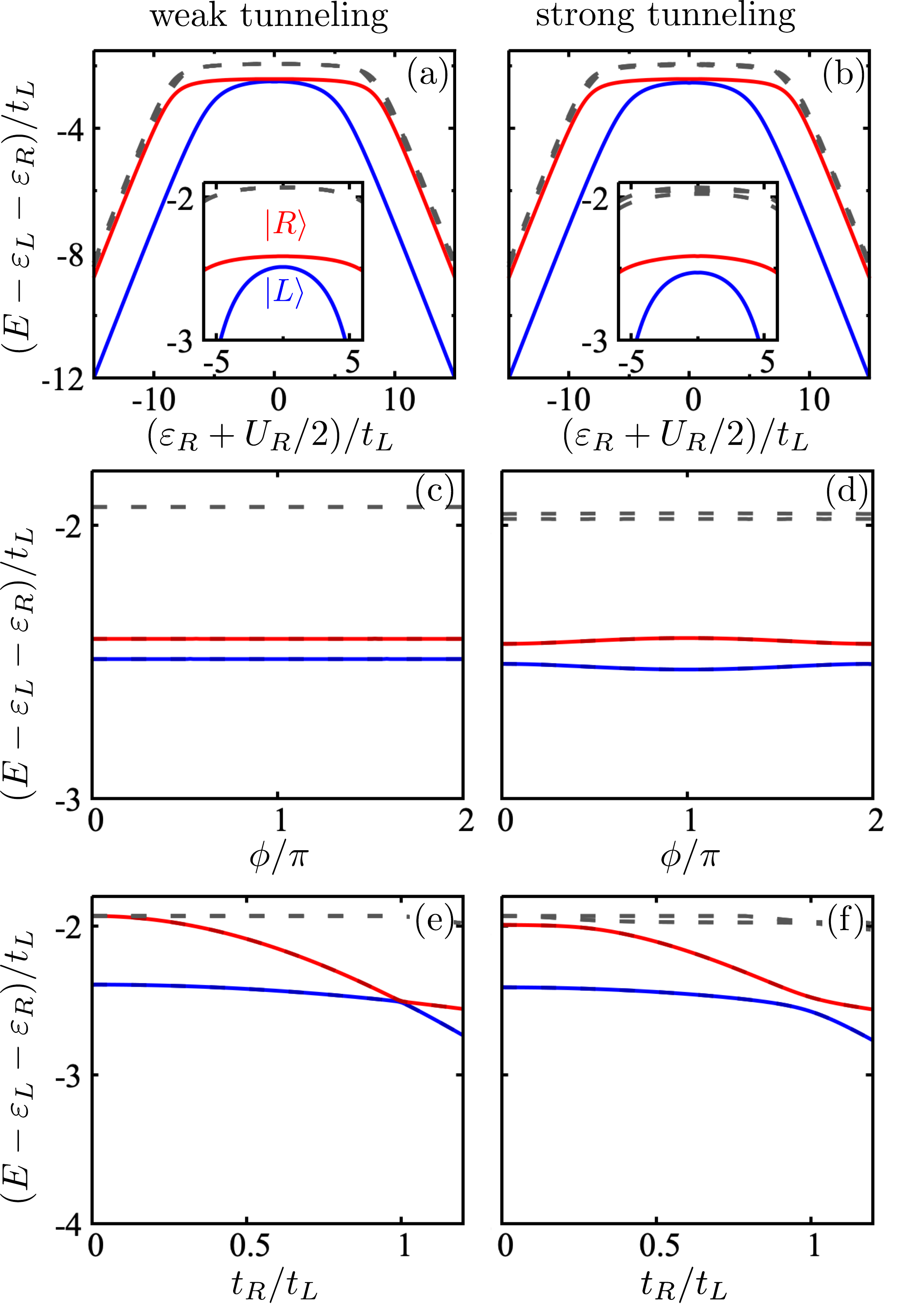}
    \caption{Spectra as a function of level energy $\varepsilon_\nu$, phase difference $\phi$, and dot-superconductor tunneling amplitude $t_R$ for the weak (left) and strong tunneling regime (right). The spectra are calculated for $U \gg \Delta$ in zero-bandwidth approximation. Here, the parameters are the same as in Fig.~\ref{fig:ZBA_tau} and using $\tau = 0.05 t_L$ for weak and $\tau = 0.5 t_L$ for the strong tunneling regime.}
    \label{fig:ZBA_eps_phi_t_Bz}
\end{figure}

{\em Level energy $\varepsilon_\nu$.---} 
The spectra as a function of level energy $\varepsilon_\nu$ are shown in Fig.~\ref{fig:ZBA_eps_phi_t_Bz} (a) and (b). 
Comparing these results with the calculations in the infinite gap limit (Fig.~\ref{fig:1}), we find that the dependence of the energy difference between ground and excited states on $\varepsilon_\nu$ is much weaker than for the infinite gap limit. We attribute this to the formation of Yu-Shiba-Rusinov singlets which have a mean occupation close to one electron on the quantum dot in the range $\varepsilon_\nu \in [-U_\nu,0]$, which is much larger than the range given by $\Gamma_\nu$ obtained for the infinite gap limit (Fig.~\ref{fig:1}).

{\em Phase difference.---}
As a function of phase difference $\phi$ [Fig.~\ref{fig:ZBA_eps_phi_t_Bz} (c) and (d)], the weak tunneling regime exhibits a flat dispersion while for strong tunneling there is a quadratic sweet spot around $\phi = \pi$. The flat dispersion for weak tunneling occurs due to the suppression of tunneling between $|L\rangle$ and $|R\rangle$ as discussed in the paragraph on the interdot tunneling $\tau$ above.

{\em Dot-superconductor tunneling $t_R$.---}
The dot-superconductor tunneling strength $t_R/t_L$ [Fig.~\ref{fig:ZBA_eps_phi_t_Bz} (e) and (f)] tunes the asymmetry of the two quantum dots and thereby the tunneling regime. The anticrossing around the degeneracy $t_L = t_R$ is determined by the interdot tunneling $\tau$ that is suppressed by the formation of Yu-Shiba-Rusinov singlet with the superconductor (see paragraph on interdot tunneling strength $\tau$ above). 

{\em Considerations for read-out, initialization, and gates.---} 
We expect that also for the strong interaction case, the fermion parity can be initialized using the singlet-doublet transition that has been demonstrated experimentally in this regime \cite{BargerbosPRXQuantum2022Jul}. The scheme requires detuning $\varepsilon$ on a range of $U/2$, at which the local even- and odd-parity states also are distinguishable by the charge on the quantum dot. This suggests that the qubit states can be read out using charge measurements as described in Sec.~\ref{sec:qubit_initialization}. Alternatively, by detuning the phase difference $\phi$ away from the operating point $\phi=\pi$, the qubit ground and excited state have opposite supercurrents. Thus, the flux through the loop depends on the qubit state which can be measured by inductive coupling. This signal would be proportional to the inter-dot tunneling amplitude $\tau$ which favors tuning the system to the strong tunneling regime for readout.
We expect that single-qubit gates can be performed electrically by driving $\tau$ and $\varepsilon_\nu$ as in Sec.~\ref{sec:qubit_rot} even though larger amplitudes may be required. The considerations for two-qubit gates from Sec.~\ref{sec:qubit_twoqubitgates} also apply here.

\section{Gate times and leakage out of the computational subspace}
\label{sec:gate_times}

In case the Hamiltonian, driving, and noise terms respect a SU(2) spin-rotation symmetry, the qubit states are spin-degenerate.
Gates can be performed within time scales of the order of the energy separation to the next excited states $\approx h/\Gamma_\nu \ll 2\pi/\omega_0$, which is fast compared to the qubit frequency $2\pi / \omega_0$.

If SU(2) spin-rotation symmetry is broken (e.g., due to spin-orbit coupling,
\cite{FlindtPhysRevLett2006Dec,BulaevPhysRevLett2007Feb},  Zeeman coupling, or Hyperfine interaction), uncontrolled rotation of the spin leads to dephasing and compromise gate fidelity. This can be avoided by applying an external field to spin-polarize the qubit states. Then, gate times are limited by the splitting to the next excited states.

We estimate the feasibility of this procedure for two candidate semiconductor materials. The III/V semiconductor InAs has an approximately isotropic g-factor of $g_{\rm InAs} \approx -5$ to $-15$ depending on experimental details \cite{PakmehrApplPhysLett2015Aug,IrieApplPhysExpress2019May}. This $g$-factor is much larger than the g-factor $g_{\rm Al} \approx 2$ in Al superconductor \cite{vanWeerdenburgSciAdv2023Mar}. Thin and narrow Al superconducting strips can withstand in-plane magnetic fields approaching the Clogston-Chandrasekhar limit of a few T \cite{vanWeerdenburgSciAdv2023Mar} depending on thickness, which allows the quantum dots to experience for Zeeman energies $\frac{1}{2} g_{\rm InAs} \mu_B B$ exceeding the superconducting gap $\Delta \approx 0.2\ $meV to $0.5\ $meV in Al thin films \cite{vanWeerdenburgSciAdv2023Mar}. 
In superconductor-semiconductor heterostructures, the induced pairing gap in the presence of an in-plane field is further limited by the orbital effect giving rise to a Cooper pair momentum in the semiconductor \cite{BanerjeePhysRevLett2023Nov}. For a thin and narrow superconductor, the induced gap closes when one flux quantum encloses the area given by the strip width and center-to-center distance $d$ between superconductor and two-dimensional electron gas. For a typical width of $W = 200$~nm, the orbital-induced gap closure occurs around $\Phi_0 / d W \approx 700$~mT where we used $d = 15~$nm as a typical value for InAs/Al heterostructures \cite{BanerjeePhysRevLett2023Nov,HaxellACSNano2023Sep}.
Already $B = 200$~mT with $g_{\rm InAs} = -10$ achieves a Zeeman splitting $g_{\rm InAs} \mu_B B = 0.12\ {\rm meV} = 2\pi\hbar\times 30 \ {\rm GHz}$ of the order of $\Delta$. We expect this to allow gate times on the nanosecond scale. Such fast gate times are typical for two-level systems well separated from the excited states, such as in spin qubits \cite{Hays_Science2021,BurkardRevModPhys2023Jun,PitaVidalNatPhys2023Aug}.
We expect this procedure also to apply for other III/V semiconductors with sufficiently large $g$-factors.

As a second example, we consider the group-IV semiconductor SiGe. This material has a strongly anisotropic g-factor, with in-plane $g_{{\rm SiGe};\parallel} \approx -0.3$ and out-of-plane g-factor $g_{{\rm SiGe}; \perp} \approx - 15$ \cite{MillerPhysRevB2022Sep}. SiGe makes good interfaces with Al as parent superconductor \cite{Scappucci_NRM2021,vanWeerdenburgSciAdv2023Mar}. The out-of-plane critical field of Al is around $100~$mT \cite{vanWeerdenburgSciAdv2023Mar,Torras-ColomaarXiv2023Aug} and may further be improved by nitridization \cite{Torras-ColomaarXiv2023Aug}. 
Taking $40$~mT as a reasonable value where a hard superconducting gap is still observed, Zeeman splittings of the order of $g_{{\rm SiGe} \perp} \mu_B B \approx 0.035\ {\rm meV} \approx 2\pi\hbar\times 8 \ {\rm GHz}$ permit gate times on the nanosecond scale. 
It may further be possible to enhance the critical out-of-plane magnetic field by using thin and narrow strips where vortex formation (which suppresses superconductivity in the type-1 superconductor Al) is avoided up to magnetic fields $B_{\rm c} \approx \Phi_0 / W^2$ with the film width $W$ \cite{StanPhysRevLett2004Mar}. For a typical width $W = 100$~nm, we have $B_{\rm c} \approx 200$~mT.

Finally, to assess gate fidelity, we note that for single-qubit gates by driving $\varepsilon_\nu$, it is not necessary to detune the system away from the charge sweet spots, {\it c.f.} App.~\ref{app:Rabi_sweet_spot}. 
Furthermore, in case the closest excited states are of opposite spin, electrical driving schemes couple to these states only indirectly via spin-orbit coupling and similar effects which may yield only weak leakage into these states.
Performing two-qubit gates requires to detune $\varepsilon_\nu$ (for capacitive coupling) or tune to the strong tunneling regime (for inductive coupling). While performing the gate the system is thus sensitive to charge noise coupling to $\varepsilon_\nu$ or flux noise $\phi$, respectively. Optimal operation will further depend on the achievable mutual inductance and charging energy.

\section{Estimate of coherence times}
\label{sec:coherence_times}

We estimate the most relevant decoherence sources in two semiconductor platforms: the group III/V semiconductor InAs and the group IV semiconductor platform SiGe. 

In InAs, Overhauser exchange fields couple the nuclear spin bath to the spin of the quantum dots. This limits the dephasing time of spin qubits in this platform to tens of nanoseconds \cite{Hays_Science2021,PitaVidalNatPhys2023Aug}. In our platform, a large magnetic field can be applied to polarize the spin in both quantum dots (see Sec.~\ref{sec:gate_times}). Then, only Overhauser fluctuations along the direction of the applied field lead to dephasing. Operating our qubit in the weak tunneling regime which is linearly sensitive to Overhauser fields and assuming Overhauser field as the dominant dephasing mechanism, we expect an improvement in dephasing time by a factor of 3 compared to spin qubits in this platform \cite{Hays_Science2021,PitaVidalNatPhys2023Aug} due to the suppression of fluctuations perpendicular to the applied field. The strong tunneling regime has a quadratic insensitivity to Overhauser fluctuations but is linearly sensitive to electric field noise coupling to the interdot tunneling $\tau$. 
The noise acting on interdot tunneling depends on precise device parameters, such as the distance of the gates to the system. 
A previous study on semiconductor charge qubits in GaAs \cite{PeterssonPhysRevLett2010Dec}, the coherence time at the charge sweet spot $\tau > |\varepsilon_L - \varepsilon_R|$ was limited to below 10~ns. This is of order of the charge relaxation time in the setup, so that the limiting mechanism could not be conclusively identified. 
In contrast, a recent work implementing a charge qubit using solid neon as host material found large coherence time of around $100\ \mu$s \cite{ZhouNatPhys2023Oct}. This suggests that the decoherence due to electric field coupling to the interdot tunneling strongly depends on material and device details. 
Altogether, based on the previous experimental results, we expect that a realization of the parity qubit in InAs/Al can improve over the realized Ramsey coherence times of $10\,$ns \cite{PeterssonPhysRevLett2010Dec}  for strong tunneling and $11\,$ns \cite{PitaVidalNatPhys2023Aug} and $18\,$ns \cite{Hays_Science2021}, proving a lower bound for the expected coherence times. 

Proximity-induced superconductivity in group IV semiconductors has been demonstrated in Ge/SiGe with Al \cite{Scappucci_NRM2021} or PtSiGe \cite{Tosato2023Apr}. In Ge/SiGe, Ramsey dephasing times in hole spin qubits of $84\,$ns \cite{Wang2022Jan} and in singlet-triplet qubits of $1\,\mu$s \cite{Jirovec2021Aug} have been demonstrated.
This platform could thus be operated in the weak tunneling regime where the qubit is quadratically insensitive to electric field noise coupling to both $\varepsilon_\nu$ and $\tau$. 
In this case, noise coupling to the tunneling rate $\Gamma_\nu$ between dot and superconductor may be a relevant factor. Our proposal may exhibit better coherence than spin qubits in this platform due to the relaxation protection by the spatial separation of the qubit states as well as charge sweet spots. 

Finally, in group IV semiconductors, dephasing due to coupling to spinful nuclei can be reduced by isotopic purification. In isotopically purified SiGe, a recent study has shown dephasing times of spin qubits of $28 \mu$s \cite{StrucknpjQuantumInf2020May}. It has been argued that the dephasing time is not limited by Overhauser coupling to the remaining spinful nuclei, but instead by coupling to electric fields and magnetic field gradients. 

\section{Comparison to other qubit realizations}
\label{sec:intro_qubits}

This section briefly summarizes how the parity qubit compares to other mesoscopic solid-state qubits in terms of principle and expected coherence properties.

{\em Semiconductor charge qubit.---} Our proposal resembles
the semiconductor charge qubit \cite{HayashiPhysRevLett2003Nov,FujisawaPhysicaE2004Mar,PeterssonPhysRevLett2010Dec} 
where the position of a single excess electron defines the qubit states. The dephasing times of charge qubits is typically in the nanosecond range \cite{HayashiPhysRevLett2003Nov,PeterssonPhysRevLett2010Dec}, 
which is attributed to coupling of the charge dipole moment of the qubit states to the environment. The parity qubit also uses the fermion occupation as information carrier, but unlike the charge qubit, the two computational states 
are both in a sweet-spot configuration which suppresses the coupling due to electric field fluctuations. Moreover, the fermion-parity qubit can achieve quadratic protection against electric field noise in both dot potential and (in the weak-tunneling regime) interdot tunneling.

{\em Semiconductor spin qubit.---}
Another related qubit is spin qubits \cite{BurkardRevModPhys2023Jun}. For III-V semiconductor realizations, the spin qubits suffer from dephasing due to fluctuations in the host nuclei spins. This will also be the case of for our parity qubit, unless (i) a large magnetic field is applied or (ii) by increasing the tunnel coupling of the quantum dots such that the qubit states are close-to-equal superpositions of $|L\rangle$ and $|R\rangle$ [see discussion around Eqs.~\eqref{eq:omega01^(2)_Bz} and \eqref{eq:omega01^(2)_Bx}]. A more promising route is silicon or germanium based systems \cite{ZwerverNatElectron2022Mar, GilbertNatNanotechnol2023Feb, Scappucci_NRM2021}. Gates in spin qubits are typically performed via electric fields taking advantage of either magnetic field gradients or spin-orbit coupling \cite{BurkardRevModPhys2023Jun,GilbertNatNanotechnol2023Feb}. The parity qubit couples in this sense more directly to electric because the qubit encoding is spatial.

{\em Andreev and Andreev spin qubits.---}
In superconductors, coherent control of single fermionic quasiparticles has been demonstrated in Andreev qubits where the computational states are defined by the occupation of spin-degenerate Andreev bound states~ \cite{Janvier_Science2015,Hays_PRL2018}. 
The Andreev qubit experiment in Al break junctions \cite{Janvier_Science2015} found Ramsey coherence times of around 38~ns, while in InAs \cite{Hays_PRL2018} the Ramsey coherence times were immeasurably short. The main decoherence sources were attributed to gate fluctuations. We expect that sweet-spot engineering in the parity qubit can protect against this dephasing source.

Further experiments on Andreev spin qubits demonstrated coherent manipulation of the quasiparticle spin ~\cite{Hays_Science2021,PitaVidalNatPhys2023Aug}. These experiments were done in Al-based structures with relatively long parity and $T_1$ lifetimes, but with somewhat shorter the dephasing times \cite{Janvier_Science2015, Hays_PRL2018, Hays_Science2021,BargerbosPRXQuantum2022Jul}. 
These were attributed to low-frequency gate-potential fluctuations \cite{Janvier_Science2015,Hays_PRL2018}. For Andreev spin qubits the nuclear spin bath was similarly to usual spin qubits suggested to be the dominant dephasing mechanism \cite{Hays_Science2021,PitaVidalNatPhys2023Aug}.
We expect that the parity qubit can mitigate fluctuations coupling to spin (Overhauser, magnetic fields, spin-orbit) because it is compatible with applying an external in-plane Zeeman field of a few 100~mT polarizing the spin (see also Sec.~\ref{sec:gate_times}). This suppresses dephasing because only fluctuations along the applied field contribute. By choosing either the weak or strong tunneling regime, the coherence can be optimized depending on the relative impact of fluctuations coupled to the spin degree of freedom and electric field noise coupling to $\tau$. 

{\em Pairs of quantum dots coupled to superconductors.---}
Recently, other proposals for qubits in pairs of quantum dots coupled to superconductors have appeared. Ref.~\onlinecite{MalinowskiarXiv2023Mar, Gorm2023} discusses two quantum dots coupled to a single superconducting island. In these proposals, similar charge-insensitive sweet spots can be reached. 

{\em Majorana qubit.---}
Our proposal shares two essential properties with topological Majorana qubits \cite{LutchynNatRevMater2018May, PradaNatRevPhys2020Oct,FlensbergOppenStern2021}, namely that the information is encoded in the fermion parity and thus separated from electronic charge. However, there are fundamental differences. In Majorana qubits, the quantum information is stored in spatially separated Majorana bound states, where a pair of Majorana bound states composes a single fermion. Dephasing of quantum information stored in Majorana bound states due to local fluctuations occurs only via modifications of the hybridization between Majorana bound states (which, in the topological case, is exponentially suppressed). In contrast, in our proposal the quantum dots are fine-tuned to a sweet spot with zero charge dipole moment. In spirit, our proposal is thus related to two-dot Kitaev chain proposals \cite{LeijnseFlensberg2012, SauNatCommun2012Jul, TomDvirNature2023Feb}. However, this so-called "poor man's" Majorana system has qubit states defined by the total parity of two dots (and therefore needs four dots to make a workable qubit), whereas for our proposal the qubit states are defined by local parity (and therefore two dots are sufficient). 
Moreover, we note that while it remains challenging to provide convincing experimental evidence of coherent manipulation of Majorana zero modes, the parity qubit is based on established technology ({\it e.g.} in InAs/Al: hybridization \cite{WangNature2022Dec} and coherent control of Andreev bound states \cite{Janvier_Science2015, Hays_PRL2018, Hays_Science2021, PitaVidalNatPhys2023Aug}, fermion-parity initialization \cite{WesdorparXiv2021Dec, BargerbosPRXQuantum2022Jul} and read-out \cite{vanDrielarXiv2023Nov}, and integration with superconducting circuits \cite{BargerbosPRXQuantum2022Jul}).

In comparison to poor man's Majorana systems, our proposal requires less technological overhead as only two instead of four dots \cite{TsintzisarXiv2023Jun,PinoarXiv2023Sep,SamuelsonarXiv2023Oct} are required to define a qubit (see above). Poor man's Majorana systems have perfect protection from fluctuations of $\varepsilon_\nu$ in only one of the dots as long as the other remains at the sweet spot, but only have quadratic protection from common-mode fluctuations $\varepsilon_L = \varepsilon_R$ \cite{LeijnseFlensberg2012}. Furthermore, poor man's Majorana systems are linear sensitive to fluctuations in the interdot-tunneling amplitude $\tau$ and the induced superconducting correlations $\Gamma$ \cite{LeijnseFlensberg2012}. A recent proposal has suggested that a poor man's Majorana system can also be realized in the same double-quantum dot structure considered in our work \cite{SamuelsonarXiv2023Oct}.

{\em Superconducting qubits.---}
Superconducting qubits (such as transmons \cite{KjaergaardAnnuRevCondensMatterPhys2020Mar} and fluxoniums \cite{ManucharyanScience2009Oct,EarnestPhysRevLett2018Apr,NguyenPhysRevX2019Nov}) are currently among the leading qubit candidates due to their long coherence times and well-developed technology. They are very different in nature from the parity qubit and other mesoscopic qubits mentioned above. One difference is the size of the qubits which for the parity qubit is determined by the size of the loop, which is much smaller than the superconducting qubits. Another difference is that the limiting factors are to a large extend determined by the more macroscopic device layout factor, whereas the mesoscopic qubits are limited by fabrication, tuning, and material issues, as mentioned above.

\section{Conclusions and discussions}
\label{sec:conclusions}
We have proposed a qubit where quantum information is encoded in the fermion parity of two quantum dots constituting a weak link in a superconducting loop.
The qubit states are defined by the fermion number parity of the two quantum dots. By electrostatic tuning to a sweet spot, the quantum dots have the same electric charge, independent of their fermion parity.
Thereby, the qubit states are protected from dephasing to first order in fluctuations of the electric field that lead to variations of the quantum dot level energy much smaller than the induced superconducting correlations 
(for infinite gap $\Delta \gg U_\nu, \Gamma_\nu$, {\it c.f.} Fig.~\ref{fig:1}) or the charging energy (for large $U_\nu \gg \Delta, \Gamma_\nu$, {\it c.f.} Fig.~\ref{fig:ZBA_eps_phi_t_Bz}).
By tuning the tunneling strength between the quantum dots, the encoded quantum information is further protected against relaxation (weak tunneling) or dephasing (strong tunneling) induced by environmental fluctuations coupling to individual quantum dots, including electric and magnetic fields, and nuclear spins. 
For weak tunneling, the qubit also exhibits a quadratic sweet spot protecting against fluctuations in the interdot tunneling strength.
The reduced sensitivity towards noise acting on indivial quantum dots may mediate the main decoherence sources in qubits in quantum dot-superconductor heterostructures, such as Andreev \cite{Janvier_Science2015,Hays_PRL2018} and Andreev spin qubits \cite{Hays_Science2021,PitaVidalNatPhys2023Aug} [see Sec.~\ref{sec:intro_qubits}]. 
Strong tunneling increases the dephasing due to fluctuations of the tunneling amplitude, which could be reduced by lowering the lever-arm between gate voltage and tunneling strength.
In systems without spin-splitting, the fermion-parity qubit can be operated purely electrically. If spin-splitting by effects such as spin-orbit coupling, external magnetic fields, or hyperfine coupling to nuclear spins is relevant, the qubit subspace can be spin-projected by an external applied magnet field which reduce the impact of noise due to fluctuations coupling to spin.

Single- and two-qubit gates, as well as initialization and readout can be performed using direct coupling to electric gates.
The qubit states are well separated from the nearest non-computational states by an energy difference of the order of the induced pairing potential $2\Gamma_\nu$, which permits fast gate operations while avoiding population of non-computational states, similar to spin qubits. 
In semiconductor platforms InAs and SiGe with Al as parent superconductor, we estimated that single-qubit gates can be performed on a time scale of 1~ns. We expect that the fermion parity qubit can achieve reasonable coherence in platforms with either low noise on the interdot tunneling strength or low Overhauser spin-spin exchange noise with nuclear spin, such as in group IV semiconductors where SiGe is a promising candidate due to the available technology. For the former, strong tunneling protects towards Overhauser noise. For the latter, operating the system in the weak-tunneling regime protects against electric field noise coupling to both inter-dot tunneling and level energies.

Future theoretical directions could further investigate how correlation effects affect and could be engineered to optimize coherence. For example, in the Yu-Shiba-Rusinov regime, the case when the interdot tunneling becomes larger than the tunneling to the superconductors could be explored. Also, it would be interesting to discuss qubit operation with quantum dots in the Kondo regime. Further directions could quantify how inelastic scattering with phonons or quasiparticles affect coherence. 

An experimental realization of our proposal would demonstrate encoding of quantum information in the fermionic parity degree of freedom separated from its electric charge. This property is shared by the topological Majorana qubits and is an essential element in their anticipated decoupling of the encoded quantum information from the environmental noise. While topological Majorana qubits \cite{LutchynNatRevMater2018May, PradaNatRevPhys2020Oct,FlensbergOppenStern2021} have proven difficult to realize, our proposal uses currently available technology. We hope our proposal inspires experimental realization and further studies on how superconductivity can decouple quantum information from the environment.

\section{Acknowledgements}
We thank Anasua Chatterjee, Ferdinand Kuemmeth, Charles M. Marcus, Jens Paaske, and Gorm Ole Steffensen for discussions. We acknowledge support from the Danish National Research Foundation, the Danish Council for Independent Research \textbar \ Natural Sciences, the European Research Council (Grant Agreement No. 856526), Spanish CM “Talento Program” (project No. 2022-T1/IND-24070), the Swedish Research Council (Grant Agreement No. 2020-03412), NanoLund, the Spanish Ministry of Science, innovation, and Universities through Grant PID2022-140552NA-I00, and the German Research Foundation under the Walter Benjamin program (Grant Agreement No. 526129603). This project has received funding from the European Union’s Horizon 2020 research and innovation program under the Marie Sklodowska-Curie grant agreements No. 101034324 and No. 101063135.

\appendix

\section{Rabi driving at the sweet spot}
\label{app:Rabi_sweet_spot}
Rabi transitions of the qubit states can also be achieved by driving the level energy $\varepsilon_\nu$ at the sweet spot. In this regime, we have to account for the quadratic dependence of the qubit frequency on the detuning of the level energy around the sweet spot, $\hbar \omega_{0}(\varepsilon_\nu) = \hbar \omega_{0}(-U_\nu/2) + \frac{1}{2}\frac{\partial^2 \omega_0}{\partial \varepsilon_\nu^2} (\varepsilon_\nu + U_\nu/2)^2 + \mathcal{O}\left((\varepsilon_\nu + U_\nu/2)^4\right)$, where the factor $\frac{1}{2}\frac{\partial^2 \omega_0}{\partial \varepsilon_\nu^2}$ can be determined from perturbation theory, see App.~\ref{app:decoherence} for an explicit expression. 
Due to the quadratic dependence of the low-energy Hamiltonian projected on the qubit subspace in the driven parameter $\varepsilon_\nu(t) = -U_\nu/2 + \delta \varepsilon_\nu \cos(\Omega t)$, the drive enters the qubit subspace as $\frac{1}{2}\frac{\partial^2 \omega_0}{\partial \varepsilon_\nu^2} (\varepsilon_\nu(t) + U_\nu/2)^2 = \frac{1}{4}\frac{\partial^2 \omega_0}{\partial \varepsilon_\nu^2} \delta \varepsilon_\nu^2 [1 + \cos(2 \Omega t)]$ with doubled frequency. 
For small driving amplitudes, $\frac{1}{4}\frac{\partial^2 \omega_0}{\partial \varepsilon_\nu^2} \delta \varepsilon_\nu^2 \ll \hbar \omega_0$, the driving frequency $\Omega$ needs to be set to half of the qubit frequency at the sweet spot, $\Omega = \omega_{0}/2$ to achieve complete population transfer. For finite driving amplitudes, $\frac{1}{4}\frac{\partial^2 \omega_0}{\partial \varepsilon_\nu^2} \delta \varepsilon_\nu^2 \gtrsim \hbar \omega_0$, the qubit frequency $\tilde{\omega}_{0} = \frac{1}{T}\int_0^T dt\, \omega_{0}[\varepsilon_\nu(t)] = \hbar \omega_0(-U_\nu/2) + \frac{1}{4}\frac{\partial^2 \omega_0}{\partial \varepsilon_\nu^2} \delta \varepsilon_\nu^2$ averaged over one period of the drive $T = \frac{2 \pi}{\Omega}$ deviates from the non-driven value $\omega_{0}$ such that the driving frequency needs to be corrected as $\Omega = \frac{1}{2}\tilde{\omega}_{0} = \frac{1}{2} \omega_0 + \frac{1}{8}\frac{\partial^2 \omega_0}{\partial \varepsilon_\nu^2} \delta \varepsilon_\nu^2$ to allow complete population transfer. The sweet spot driving is demonstrated in Fig.~\ref{fig:3}(b). A similar result also allows driving of the phase difference $\phi$ at the sweet spot $\phi = \pi$ to perform qubit rotations. 

\section{Perturbation theory and decoherence rates}
\label{app:decoherence}

Here we give the full result of the second-order perturbation theory for the qubit frequency around the sweet spot $\varepsilon_\nu = - U_\nu/2$ and at the operating point $\phi = \pi$. These results are employed to calculate the Bloch-Redfield dephasing rates and relaxation rates from Eqs.~\eqref{eq:def_dephasing_rate} and \eqref{eq:def_relaxation_rate}.

\subsection{Diagonalization at the sweet spot}
\label{app:diagonalziation}

The full qubit Hamiltonian including spin-orbit coupling and Zeeman fields, 
\begin{equation}
\hat{H} = \sum_\nu (\hat{H}_\nu + \hat{H}^B_\nu) + \hat{H}_T^\text{soc}\,,
\end{equation}
is conveniently diagonalized in the eigenbasis of proximitized dots $\hat{H}_\nu$, 
\begin{equation}
|\lambda_\nu\rangle_\nu = \frac{|\rm{vac}\rangle  + (\sinh{(\xi_{\nu})} + {\rm s}_{\lambda_\nu} \cosh{(\xi_{\nu})}) c^\dagger_{\uparrow \nu}c^\dagger_{\downarrow \nu}|\rm{vac}\rangle}{N_{\lambda_{\nu}}} 
\label{eq:basis_Hnu}
\end{equation}
with labels $\lambda_\nu = \lambdap, \lambdam$ as in Sec.~\ref{sec:qubit_regimes}, $N_{\lambda_\nu} = \sqrt{2 \cosh{\xi_{\nu}} (\cosh{\xi_{\nu}} + {\rm s}_{\lambda_\nu} \sinh{\xi_{\nu}})}$, $\sinh{\xi_{\nu}} = \frac{2\varepsilon_{\nu}+U_{\nu}}{2\Gamma_{\nu}}$ and energy $E_{\lambda_\nu} = \Gamma_\nu(\sinh{\xi_{\nu}} + {\rm s}_{\lambda_\nu} \cosh{\xi_{\nu}})$. 

In the eigenbasis of proximitized dots $\hat{H}_\nu$ as given in Eq.~\eqref{eq:basis_Hnu}, the Hamiltonian within a spin sector $\sigma$ can be written in matrix form,
\begin{equation}
\hat{H}_\sigma = \begin{pmatrix}
\varepsilon_{\sigma L}+E_{\lambdap_R} & 0 & \tau e^{i \sigma \theta} f_{\lambdap \lambdap} & \tau e^{i \sigma \theta}f_{\lambdap \lambdam}\\
0 & \varepsilon_{\sigma L}+E_{\lambdam_R} & \tau e^{i \sigma \theta}f_{\lambdam \lambdap} & \tau e^{i \sigma \theta}f_{\lambdam \lambdam}\\
&  & \varepsilon_{\sigma R}+E_{\lambdap_L} & 0\\
\text{H.c.} &  & 0 & \varepsilon_{\sigma R}+E_{\lambdam_L}
\end{pmatrix}
\label{eq:full_Hamiltonian_basis_Hnu}
\end{equation}
acting on the basis $(\hat{c}^\dagger_{\sigma L}|\lambdap\rangle_R, \hat{c}^\dagger_{\sigma L}|\lambdam\rangle_R, \hat{c}^\dagger_{\sigma R}|\lambdap\rangle_L, \hat{c}^\dagger_{\sigma R}|\lambdam\rangle_L)^{\rm T}$, where $\varepsilon_{\sigma \nu} = \varepsilon_\nu + {\rm s}_\sigma B_{z, \nu}$ and
\begin{align}
f_{\lambda_R\lambda_L} &=\frac{e^{i\phi/2}}{N_{\lambda_{R}}N_{\lambda_{L}}} \label{eq:f_basis_Hnu} \\ \nonumber
 & \ - \frac{{\rm s}_{\lambda_{R}}{\rm s}_{\lambda_{L}}e^{-i\phi/2}}{N_{\lambda_{R}}N_{\lambda_{L}}} \big( \cosh({\rm s}_{\lambda_{L}}\xi_{L}+{\rm s}_{\lambda_{R}}\xi_{R}) \\ \nonumber
& \qquad \qquad \qquad \quad \ +\sinh({\rm s}_{\lambda_{L}}\xi_{L}+{\rm s}_{\lambda_{R}}\xi_{R}) \big)  \nonumber
\end{align}
At the sweet spot $\xi_L = \xi_R = 0$, the function $f_{\lambda_R\lambda_L}$ simplifies as $f^{(0)}_{\lambda\lambda}=i \sin{\frac{\phi}{2}}$ and $f^{(0)}_{\lambda\bar{\lambda}}=\cos{\frac{\phi}{2}}$ where $\bar{\lambdap} / \bar{\lambdam} = \lambdamp$. Thus, the tunnel-coupling of the quantum dot states can be controlled by the phase-bias $\phi$: At $\phi = 0$, the low-energy state $|\lambdam\rangle_L$ of the left dot is coupled to the high-energy state $|\lambdap\rangle_R$ of the right dot, and vice versa. At $\phi = \pi$, the low-energy states $|\lambdam\rangle_L$, $|\lambdam\rangle_R$ and high-energy states $|\lambdap\rangle_L$, $|\lambdap\rangle_R$ are coupled with each other. The perpendicular components of the Zeeman field $B_{x, \nu},\ B_{y, \nu}$ remain unaffected by this basis transformation.

At the sweet spot, for $B_{x, \nu} = B_{y, \nu} = 0$, and at $\phi = \pi$, the eigen states of the full Hamiltonian $\hat{H}$ are,
\begin{align}
|\sigma,\lambda,\rho=0\rangle & = \sin{\frac{\eta_{\sigma,\lambda}}{2}} \hat{c}^\dagger_{\sigma L}|\lambda\rangle_R + i e^{-i {\rm s}_\sigma \theta} \cos{\frac{\eta_{\sigma,\lambda}}{2}} \hat{c}^\dagger_{\sigma R}|\lambda\rangle_L \nonumber \\
|\sigma,\lambda,\rho=1\rangle & = \cos{\frac{\eta_{\sigma,\lambda}}{2}} \hat{c}^\dagger_{\sigma L}|\lambda\rangle_R - i e^{-i {\rm s}_\sigma \theta} \sin{\frac{\eta_{\sigma,\lambda}}{2}} \hat{c}^\dagger_{\sigma R}|\lambda\rangle_L \label{eq:eigenstates_SS_basis_Hnu}
\end{align}
with energy
\begin{align}
E_{\sigma,\lambda,\rho}^{(0)} & = \frac{-(U_R + U_L)/2 + {\rm s}_\sigma(B_{z L} + B_{z R}) + {\rm s}_\lambda(\Gamma_R + \Gamma_L)}{2} \nonumber \\ 
& \ + {\rm s}_{\rho} \frac{\tau}{\sin{\eta_{\sigma,\lambda}}},
\label{eq:energy_SS_basis_Hnu}
\end{align}
with $\rho = 0, 1$ labeling the corresponding eigen state and the superscript $(0)$ indicates the sweet spot. The angle 
$\eta_{\sigma, \lambda}$ is given in Eq.~\eqref{eq:eta_SS_basis_Hnu}. It determines the operating regime: For small $|\eta_{\sigma,\lambdam}| \ll \pi/2$ (weak tunneling), the eigen states Eqs.~\eqref{eq:eigenstates_SS_basis_Hnu} are localized either left or right, while for large $|\eta_{\sigma,\lambdam}| \approx \pi/2$ (strong tunneling), the eigen states are bonding and anti-bonding superpositions between left and right. 

The two qubit states are given by the two lowest energy eigen states of the system within a spin sector $\sigma$, $|\sigma,\lambdam,-\rangle$ and $|\sigma,\lambdam,+\rangle$. 
The qubit eigenfrequency $\hbar \omega_{0}^{(0)} = E^{(0)}_{\sigma, \lambdam, +} - E^{(0)}_{\sigma, \lambdam, -}$ is given in Eq.~\ref{eq:omega0_with_Zeeman}.
For Zeeman fields (and fluctuations thereof) much smaller than $\hbar \omega_{0}^{(0)}$, the system can be operated in the spin degenerate regime. For sizeable fluctuations of the Zeeman fields, it is advantageous to apply an external magnetic field $B^\text{ext}_{z\nu}$ much larger than the variance of the fluctuations to suppress spin flips. Differences in the electronic $g$-factor in the two quantum dots $g_\nu$ lead different resulting Zeeman fields $B^\text{ext}_{z\nu}$.

\subsection{Deviations from the operating regime}
\label{sec:noise_deviations}
Here we summarize the results for the lowest order corrections to the qubit spectrum due to perturbations in the system parameters away from the operating point at the sweet spot $\varepsilon_\nu = - U_\nu/2$ and $\phi = \pi$.

{\em Detuning $\varepsilon_{\nu}$.---} We calculate the energy difference $\hbar \omega_{0}^{(2)} = E_{\sigma,\lambdam,+}^{(2)}-E_{\sigma,\lambdam,-}^{(2)}$ of the qubit states to second order in the detuning,
\begin{align}
\label{eq:omega01^(2)_detuning_app}
\hbar \omega_{0}^{(2)}&=\hbar \omega_{0}^{(0)}
+\frac{\Gamma_{L}\xi_L^2 - \Gamma_{R}\xi_R^2}{2} \cos\eta_{\sigma,\lambdam} \\
&\ -\left(\frac{\xi_{L}-\xi_{R}}{2}\right)^{2}\left(1 + 2 \Xi_{\sigma,\lambdam}^2 \right)\tau\sin\eta_{\sigma,\lambdam} \nonumber
\end{align}
where $\xi_\nu = \text{arsinh}\frac{2\varepsilon_{\nu}+U_{\nu}}{2\Gamma_{\nu}} \approx \frac{2\varepsilon_{\nu}+U_{\nu}}{2\Gamma_{\nu}}$ and $\Xi_{\sigma,\lambda} = \frac{1}{\Gamma_R + \Gamma_L}\frac{ \tau}{\sin\eta_{\sigma,\lambda}}$. The first correction arises from the quadratic dependence of the energy $E_{\lambda_\nu}$ of the local even parity states of the individual dots. The second correction describes a modification of the tunneling between these states due to their dependence of the Andreev bound state wave function (Eq.~\eqref{eq:basis_Hnu}) on the level energy $\varepsilon_\nu$. The term proportional to $\Xi_{\sigma,\lambdam}$ describes second-order processes involving the symmetric even-fermion-parity bound states $\lambda_\nu = \lambdap$ at high energy $\lambda_\nu \Gamma_\nu$ 
at the sweet spot. The above result is furthermore expanded to lowest order in $\Xi_{\sigma,\lambda}$, which is small $\Xi_{\sigma,\lambda} \ll 1$ when the qubit states $\lambda_\nu = \lambdam$ are well separated from the high-energy states $\lambda_\nu = \lambdap$.
Neglecting the coupling to the symmetric even-fermion-parity bound states ($\Xi_{\sigma,\lambdam} \to 0$) reproduces Eq.~\eqref{eq:omega01^(2)_detuning}.

{\em Phase difference $\phi$.---} Similarly, the variations of the phase difference $\phi = \pi + \delta \phi$ away from the operating point $\phi = \pi$ modify the qubit frequency as
\begin{equation}
\hbar \omega_{0}^{(2)}=\hbar \omega_{0}^{(0)} 
-\frac{\delta\phi^2}{4}\left(1 + 2 \Xi_{\sigma,\lambdam}^2 \right)\tau\sin\eta_{\sigma,\lambdam}
\label{eq:omega01^(2)_phi_app}
\end{equation}
which, again, is expanded to second order in $\Xi_{\sigma,\lambda}$. Setting $\Xi_{\sigma,\lambdam} \to 0$ reproduces Eq.~\eqref{eq:omega01^(2)_phi}.

{\em Charging energy $U_\nu$.---} Fluctuations in the charging energy $U_\nu$ on either dot modify the qubit frequency to linear order,
\begin{equation}
\hbar \omega_{0}^{(1)}=\hbar \omega_{0}^{(0)} + \left(\frac{\delta U_{R}}{2}-\frac{\delta U_{L}}{2}\right)\cos\eta_{\sigma,\lambdam}.
\label{eq:omega01^(2)_U}
\end{equation}

{\em Induced pairing potential $\Gamma_\nu$.---} At the sweet spot, fluctuations in the induced pairing strength $\Gamma_{\nu}\to\Gamma_{\nu}+\delta\Gamma_{\nu}$ modify only the diagonal elements of the Hamiltonian. The offdiagonal terms remain unchanged as $\xi_{\nu}=0$. The energy depends linearly on the perturbation,
\begin{equation}
\hbar \omega_{0}^{(1)}=\hbar \omega_{0}^{(0)} + \left(\delta\Gamma_{L}-\delta\Gamma_{R}\right)\cos\eta_{\sigma,\lambdam}.
\label{eq:omega01^(2)_Gamma}
\end{equation}

{\em Zeeman fields $\vec{B}_\nu$.---}
Dephasing and relaxation due to Zeeman field fluctuations are discussed around Eqs.~\eqref{eq:omega01^(2)_Bz} and \eqref{eq:omega01^(2)_Bx} in the main text.

{\em Spin-orbit coupling.---} 
Spin-orbit coupling enters the Hamiltonian via the tunneling term Eq.~\eqref{eq:H_T^SOC}. Without Zeeman fields, or for Zeeman fields pointing only along the spin-orbit direction $\vec{n}$, the spin-orbit coupling does not modify the spectrum. The spin orbit angle $\theta$ enters only in the perturbative corrections due to orthogonal fluctuations of the Zeeman field [see Eq.~\eqref{eq:omega01^(2)_Bx}].

\subsection{Dephasing and relaxation}
\label{sec:noise_dephasing}
We derive Bloch-Redfield dephasing rates and relaxation rates using Eqs.~\eqref{eq:def_dephasing_rate} and \eqref{eq:def_relaxation_rate} from the main text.

{\em Detuning $\varepsilon_\nu$.---} Using Eq.~\eqref{eq:omega01^(2)_detuning}, the dephasing rate due to fluctuations in the detuning $\varepsilon_\nu$ around the sweet spot is
\begin{align}
&\frac{2 \hbar^{2}}{\pi S_{\varepsilon}(\omega\to0)} \Gamma_{\varphi}^{\varepsilon_{\nu}}= \\ \nonumber & \left(\xi_{\nu}\cos\eta_{\sigma,\lambdam}+\frac{\xi_{L}-\xi_{R}}{2\Gamma_{\nu}}\left(1 +  2\Xi_{\sigma,\lambdam}^2 \right)\tau\sin\eta_{\sigma,\lambdam}\right)^{2}.
\end{align}
To compute the relaxation rate, we calculate at the sweet spot and $\phi = \pi$
\begin{equation}
\frac{d\hat{H}}{d\varepsilon_\nu} = \openone_{4}+\frac{\tau}{2\Gamma_{\nu}}\lambda_{y}\otimes \nu_{x}
\end{equation}
where $\lambda_l$ and $\nu_l$, $l=0,x,y,z$ are Pauli matrices in $\lambda$ space and quantum dot space, respectively, and "$\otimes$" denotes the direct product. The matrix element $\langle + | \frac{d\hat{H}}{d\varepsilon_\nu} | - \rangle = 0$ due to orthogonality of the wave function and because the operator $\lambda_{y}\otimes \nu_{x}$ relates wavefunctions with opposite $\lambda$, while the two qubit states have the same $\lambda = -$. Thus we have
\begin{equation}
\Gamma_{\text{rel}}^{\varepsilon_\nu} = 0.
\end{equation}
Both the Bloch-Redfield dephasing  rate and relaxation rate are zero at the sweet spot $\xi_\nu = 0$ and $\phi = \pi$. 

{\em Phase difference $\phi$.---} The dephasing rate due to fluctuations in $\phi = \pi + \delta \phi$ is
\begin{equation}
\frac{2 \hbar^{2}}{\pi S_{\phi}(\omega\to0)}\Gamma_{\varphi}^{\phi}=\left(\frac{\delta\phi\left(1 + 2 \Xi_{\sigma,\lambdam}^2 \right)\tau\sin\eta_{\sigma,\lambdam}}{2}\right)^2.
\end{equation}
With $\frac{d\hat{H}}{d\phi} = -\lambda_x \otimes \nu_x$ at $\phi = \pi$ we find $\langle + | \frac{d\hat{H}}{d\phi} | - \rangle = 0$ because the operator $\lambda_x \otimes \nu_x$ couples only states with opposite $\lambda$, such that 
\begin{equation}
\Gamma_{\text{rel}}^{\phi} = 0.
\end{equation}

{\em Tunneling strength $\tau$.---} Fluctuations in the tunneling strength lead to a dephasing and relaxation rate,
\begin{align}
\Gamma_{\varphi}^{\tau}&=\frac{4 \pi}{\hbar^{2} }\sin^{2}\eta_{\sigma,\lambdam} S_{\tau}(\omega\to0) \\
\Gamma_{\text{rel}}^{\tau}& =\frac{\pi}{2\hbar^{2}}\cos^{2}\eta_{\sigma,\lambdam}S_{\tau}(\omega_{0})
\end{align}
where we used that $d\hat{H}/d\tau_\sigma = - \lambda_0 \otimes \nu_y$ at $\phi = \pi$.

{\em Charging energy $U_\nu$.---} Fluctuating charging energies $U_\nu$ on either dot result in dephasing and relaxation rates, to first order around the sweet spot $U_\nu = - 2 \varepsilon_\nu$,
\begin{align}
\Gamma_{\varphi}^{U_\nu}&=\frac{\pi}{4 \hbar^{2} }\cos^{2}\eta_{\sigma,\lambdam} S_{U_\nu}(\omega\to0) \\
\Gamma_{\text{rel}}^{U_\nu}& =\frac{\pi}{32 \hbar^{2}}\sin^{2}\eta_{\sigma,\lambdam}S_{U_\nu}(\omega_{0}) .
\end{align}
where we used that $d\hat{H}/dU_{L/R} = \lambda_0 \otimes (\nu_0 \pm \nu_z)/4$.

{\em Induced pairing strength $\Gamma_\nu$.---} The dephasing and relaxation rates due to fluctuating $\Gamma_\nu$
\begin{align}
\Gamma_{\varphi}^{\Gamma_\nu}&=\frac{\pi}{\hbar^{2} }\cos^{2}\eta_{\sigma,\lambdam} S_{\Gamma_\nu}(\omega\to0) \\
\Gamma_{\text{rel}}^{\Gamma_\nu}& =\frac{\pi}{8\hbar^{2}}\sin^{2}\eta_{\sigma,\lambdam}S_{\Gamma_\nu}(\omega_{0}) .
\end{align}
using $d\hat{H}/d\Gamma_{L/R} = \lambda_0 \otimes (\nu_0 \pm \nu_z)/2$.

{\em Zeeman fields $\vec{B}$.---} We again distinguish fluctuations $B_z$ along the spin quantization axis in the two dots, and perpendicular fluctuations $B_x,\ B_y$. The spin quantization axis is either set by the direction of the spin-orbit coupling $\vec{n}$ and/or the direction of an externally applied Zeeman field $B_z^\text{ext}$, see Sec.~\ref{sec:noise_deviations}. Parallel fluctuations $\delta B_z$ result in dephasing and relaxation rates 
\begin{align}
\Gamma_{\varphi}^{B_{z,\nu}}&=\frac{\pi}{\hbar^{2} }\cos^{2}\eta_{\sigma,\lambdam} S_{B_{z,\nu}}(\omega\to0) \\
\Gamma_{\text{rel}}^{B_{z,\nu}}& =\frac{\pi}{8\hbar^{2}}\sin^{2}\eta_{\sigma,\lambdam}S_{B_{z,\nu}}(\omega_{0}) .
\end{align}
In the presence of an externally applied Zeeman field much larger than the variance of the fluctuations of $B_x$ and $B_y$, fluctuations in these components yield dephasing and relaxation rates
\begin{align}
\Gamma_{\varphi}^{B_{x,\nu}}&=\frac{\pi}{\hbar^{2} }\frac{4 \delta B_{x}^{2}}{(B_{z}^{\text{ext}})^{4}} \left( \frac{\tau\sin\eta_{\sigma,\lambdam}\sin^{2}\theta}{1-(\frac{\tau}{B_{z}^{\text{ext}}\sin\eta_{\sigma,\lambdam}})^{2}} \right)^2 S_{B_{x,\nu}}(\omega\to0) \\
\Gamma_{\text{rel}}^{B_{x,\nu}}& = 0 .
\end{align}
The dephasing rate is zero in the absence of a bias $\delta B_x = 0$ and / or in the absence of spin-orbit coupling $\theta = 0$.

{\em Relative dephasing of multiple qubits.---}
When operating all qubits at the sweet spot, the charge expectation value of all quantum dots is independent of their fermion parity. Therefore, any charge dipole or multipole moments between different qubits vanish as well, such that also a multi-qubit system remains linearly insensitive to fluctuations in the level energy. Similar arguments hold for the other parameters.

\section{Pulse parameters}
\label{app:pulse_parameters}
Here, we define the precise form and parameters for the pulses applied in Fig.~\ref{fig:2} and \ref{fig:3} in the main text.

The pulses are of the form
\begin{equation}
f(t) = s(t;t_{\rm pulse}, t_{\rm rise}) \cos (\Omega t + \phi_0)
\label{eq:app_pulse_driving}
\end{equation}
with the driving frequency $\Omega$, pulse phase $\phi_0$ and envelope function 
\begin{equation}
s(t;t_{\rm pulse}, t_{\rm rise}) = \vartheta(t - t_{\rm pulse}/2; t_{\rm rise}) - \vartheta(t + t_{\rm pulse}/2; t_{\rm rise})
\label{eq:app_pulse_envelope}
\end{equation}
setting the pulse duration $t_{\rm p}$. The pulse is turned on smoothly over a rising time $t_{\rm r}$ using the smooth step function
\begin{equation}
\vartheta(t;t_{\rm rise}) = \left(1 + \exp \left[ \frac{t_{\rm rise}}{t + t_{\rm rise}/2} - \frac{t_{\rm r}}{t_{\rm rise}/2 - t} \right] \right)^{-1}
\label{eq:app_smooth_step}
\end{equation}
which is defined on the interval $-t_{\rm rise}/2 < t < t_{\rm rise}$.
This smooth step function connects continuously in derivatives of all orders to a flat line $\vartheta(t) = 0$ for $t < -t_{\rm rise}/2$ and $\vartheta(t) = 1$ for $t > t_{\rm rise}/2$.

{\em Weak tunneling, $X_\pi$ rotation [Fig.~\ref{fig:2}(b)].---}
The pulse on the tunneling amplitude is of the form $\tau(t) = \tau + \delta \tau f(t)$ with amplitude $\delta \tau = 0.006 \Gamma_L$, frequency $\Omega = \omega_0$, pulse time $t_{\rm pulse} = 115 h/\Gamma_L$ and rise time $t_{\rm rise} = 10 h/\Gamma_L$. The ramp on the level energy $\varepsilon_R$ is parameterized using the smooth step functions, $\varepsilon_R(t) = - U_R/2 + \delta \varepsilon_{R, {\rm ramp}} (1 - s(t;t_{\rm pulse} + t_{\rm ramp} + t_{\rm rise}, t_{\rm ramp}))$ with $\delta \varepsilon_{R, {\rm ramp}} = \Gamma_L$ and ramp time $t_{\rm ramp} = 30 h/\Gamma_L$. The pulse starts directly after the ramp is completed.

{\em Weak tunneling, $X_{\pi/2} Z_{2\pi} X_{\pi/2}$ sequence of rotations [Fig.~\ref{fig:2}(c)].---}
The two consecutive pulses on the tunneling amplitude are performed with $\delta \tau = 0.006 \Gamma_L$, frequency $\Omega = \omega_0$, pulse time $t_{\rm pulse} = 60 h/\Gamma_L$ and rise time $t_{\rm rise} = 10 h/\Gamma_L$. The pulses are separated by a waiting time $t_{\rm w} = 7.75 h/\Gamma_L$ and have the same phase $\phi_0$.

{\em Weak tunneling, $X_{\pi/2} Z_{2\pi} X_{\pi/2}$ sequence of rotations [Fig.~\ref{fig:2}(d)].---}
Same parameters as in Fig.~\ref{fig:2}(c), except that the phase of the second pulse is shifted $\phi_0 \to \phi_0 + \pi$.

{\em Strong tunneling, $X_{\pi}$ rotation with detuned $\varepsilon_R$ [Fig.~\ref{fig:3}(a)].---}
The pulse on the level energy $\varepsilon_R$ is applied for a detuned $\varepsilon_R(t) = -U_R/2 + \delta \varepsilon_{\rm detune} + \delta \varepsilon f(t)$, with detuning $\delta \varepsilon_{\rm detune} = 0.2\Gamma_L$, pulse amplitude $\delta \varepsilon_R = 0.1 \Gamma_L$, frequency $\Omega = \omega_0$, pulse time $t_{\rm pulse} = 55 h/\Gamma_L$ and rise time $t_{\rm rise} = 10 h/\Gamma_L$.

{\em Strong tunneling, $X_{\pi}$ rotation with by driving the level energy at the sweet spot [Fig.~\ref{fig:3}(a)].---}
The pulse to perform a $X_{\pi}$ rotation by drive the level energy at the sweet spot is of the form $\varepsilon_R(t) = -U_R/2 + \delta \varepsilon f(t)$ with amplitude $\delta \varepsilon_R = 0.3 \Gamma_L$, frequency $\Omega = \frac{1}{2} \omega_0 + \frac{1}{8}\frac{\partial^2 \omega_0}{\partial \varepsilon_\nu^2} \delta \varepsilon_\nu^2 \approx 0.500214 \omega_0$ [see App.~\ref{app:Rabi_sweet_spot}], pulse time $t_{\rm pulse} = 50 h/\Gamma_L$ and rise time $t_{\rm rise} = 10 h/\Gamma_L$.

\bibliography{bibliography}

\end{document}